\documentclass[twocolumn,showpacs,pra]{revtex4}
\usepackage{graphicx,color}
\usepackage{bm}
\usepackage{hyperref}
\usepackage{amsmath,braket}
\usepackage{amsthm,mathrsfs,amsbsy}

\newcommand{\vcompeqn}{
	\medmuskip=0mu
	\thinmuskip=0mu
	\thickmuskip=0mu}
\newcommand{\compeqn}{
	\medmuskip=1mu
	\thinmuskip=1mu
	\thickmuskip=1mu}
\newcommand{\ip}{\text{I}_\text{p}}
\newcommand{\up}{\text{U}_\text{p}}
\renewcommand\Re{\operatorname{Re}}
\renewcommand\Im{\operatorname{Im}}

\definecolor{battleshipgrey}{rgb}{0.52, 0.52, 0.51}
\definecolor{cadet}{rgb}{0.33, 0.41, 0.47}
\definecolor{charcoal}{rgb}{0.21, 0.27, 0.31}
\hypersetup{
colorlinks   = true, 
urlcolor     = battleshipgrey, 
linkcolor    = cadet, 
citecolor   = charcoal }

\begin{document}

\preprint{APS/123-QED}
\title{Coulomb-corrected quantum interference in above-threshold ionization: 
Working towards multi-trajectory electron holography}
\author{A. S. Maxwell$^1$}
\email{andrew.maxwell.14@ucl.ac.uk}
\author{A. Al-Jawahiry$^1$}
\author{T. Das$^{1,2}$}
\author{C. Figueira de Morisson Faria$^1$}
\email{c.faria@ucl.ac.uk}
\affiliation{$^1$Department of Physics \& Astronomy, University College London \\
Gower Street, London WC1E 6BT, United Kingdom\\ $^2$Max Planck Institute for Physics of Complex Systems, Dresden, N\"othnitzer Str. 38, D-01187 Dresden,
Germany}
\date{\today}

\begin{abstract}
	Using the recently developed Coulomb Quantum Orbit Strong-Field Approximation (CQSFA), we perform a systematic analysis of several features encountered in above-threshold ionization (ATI) photoelectron angle-resolved distributions (PADs), such as side lobes, and intra- and intercycle interference patterns. The latter include not only the well-known intra-cycle rings and the near-threshold fan-shaped structure, but also previously overlooked patterns. We provide a direct account of how the Coulomb potential distorts different types of interfering trajectories and changes the corresponding phase differences, and show that these patterns may be viewed as generalized holographic structures formed by up to three types of trajectories. We also derive analytical interference conditions and estimates valid in the presence or absence of the residual potential, and assess the range of validity of Coulomb-corrected interference conditions provided in the literature.
\end{abstract}


\maketitle

\section{Introduction}

Orbits and quantum interference play a vital role in phenomena that occur when matter interacts with laser fields of
intensities $I\geq 10^{14}\mathrm{W}/\mathrm{cm}^{2}$. The archetypal
description of such phenomena relies on an electron undergoing tunnel or multiphoton ionization, propagating in the continuum and either interacting
with its parent ion via recollision, or reaching the
detector directly \cite{Corkum1993}. For a given final electron momentum, there are usually many paths that
the active electron may follow. Thus, the
related probability amplitudes interfere. 

For a qualitative description of strong-field dynamics, it often suffices to neglect the binding
potential in the electron propagation and approximate the continuum by
field-dressed plane waves. This is a key idea behind the strong-field
approximation (SFA), which is one of the most widespread approaches in
strong-field and attosecond physics. Since the mid-2000s, however, many
features stemming from the interplay between the residual binding potential and the laser field have been identified in experiments.  Examples are (i) the low-energy enhancements in above-threshold ionization (ATI) spectra \cite{PRL2009Quan,NatPhys2009Blaga,PRL2010Liu,Becker2014JPB,Kaestner2012JPB,LPhys2009Shvetsov,Wu2012PRL,Pisanty2016,Dura2013SciRep,Wei2016SciR,Becker2015JPB}, (ii) the fan-shaped structure in angular resolved ATI electron momentum distributions
\cite{Rudenko2004JPB,Maharjan2006JPB,Huismanset2010Science}, and (iii) the species dependency  in nonsequential
double ionization (NSDI) with circularly polarized fields \cite{Mauger2010}.

While examples (i) and (iii) may be explained
by classical methods \cite{PRL2009Quan,PRL2010Liu,Wu2012PRL,Becker2014JPB,Kaestner2012JPB,LPhys2009Shvetsov,Wei2016SciR,Becker2015JPB,Mauger2010}, (ii) is a quantum-interference effect
that occurs near the ionization threshold. Studies of near-threshold ATI using the SFA \cite{Arbo2010PRA,ArboNuclInstr} have shown that the interference of events separated by at most half a cycle leads to nearly vertical fringes, whose distortion by the Coulomb potential leads to the fan-shaped structure.
 This relationship has been investigated 
  by modifying the final electron scattering state \cite{Chen2006PRA,ArboNuclInstr}, comparing the full solution of the time-dependent Schr\"odinger equation (TDSE) for short- and long-range potentials \cite{Chen2006PRA}, and performing classical-trajectory computations which relate the fringes to laser-dressed Kepler hyperbolae with neighboring angular momenta \cite{Arbo2006PRL,Arbo2008PRA}. One should also note that, in strong-field photodetachment of negative ions, i.e. for short-range binding potentials, there is a very good agreement between the SFA and the full solution of the time-dependent Schr\"odinger equation (TDSE) \cite{Korneev2012,Shearer2013,Hassouneh2015} and experimental results \cite{Bergues2007}, with approximately vertical fringes instead of a fan. 

In a previous publication \cite{Lai2017}, we have performed a direct analysis of how this pattern forms using the Coulomb Quantum Orbit Strong-Field Approximation (CQSFA) \cite{Lai2015}. We have shown that the fan-shaped structure may be viewed as a holographic feature caused by the interference of the trajectories that reach the detector directly, with those that are deflected by the binding potential, but do not undergo hard collisions. The phase difference between the two interfering types of trajectories is dependent on the electron scattering angle. This causes distortions in the intra-cycle fringes obtained from the SFA, which then form the fan-shaped pattern. The abovementioned work, however, left several open questions. First, in addition to the fan-shaped structure, there may be other types of intra-cycle interference, and one should clarify how the Coulomb potential distorts such patterns. Second, in photoelectron holographic structures, there are usually two types of orbits which act as probe and reference signal. Could one generalize photoelectron holography in order to incorporate additional types of orbits?

Other structures are exemplified by the ATI rings, caused by the interference of events separated by a full number of cycles, the carpet-like patterns observed in ATI angular distributions for electron emission perpendicular to the driving-field polarization \cite{Korneev2012PRL}, and a myriad of holographic structures that occur due to the interference between direct electrons and those undergoing hard collisions \cite{Huismanset2012PRL,Huismanset2010Science,Li2015SciRep,Meckel2008Science,Bian2011PRA,Bian2012PRL,Bian2014PRA,Haertelt2016PRL}. Analytic conditions have been derived for many of such structures. The overwhelming majority of these conditions, however, neglect the Coulomb potential in the electron propagation. They are either based on the SFA, or on its classical counterpart.

Nonetheless, studies employing the Coulomb corrected SFA \cite{Yan2012} show that the Coulomb potential introduces phase shifts and thus modifies interference patterns in ATI. Therein, analytic interference conditions are provided for electron emission parallel and perpendicular to the laser-field polarization, and the low-frequency limit. These conditions however are based on a series of assumptions, whose validity should be examined more closely. First, it is  postulated that only two main types of orbits contribute to the interference patterns: those that leave in the direction and from the opposite side of the detector, known as orbits type 1 and 2, respectively. This is the case in the SFA, but the Coulomb potential modifies the topology of the problem by introducing two more types of orbits \cite{Yan2010,Lai2015}. Second, one assumes that the transition amplitudes related to orbits 1 and 2 have the same absolute values for intra-cycle events, and that momenta associated with different orbits populate the same region. There is, however, no evidence that these assumptions hold in the presence of the Coulomb potential. 

In the present article, we perform a direct, quantum-orbit analysis of how the Coulomb potential influences ATI photoelectron distributions. This includes the side lobes, inter- and intra-cycle interference.  We provide analytic estimates for interference maxima and minima, and investigate to which extent the assumptions in Ref.~\cite{Yan2012} hold.  We also assess how specific patterns form, and whether one must go beyond only two types of orbits. 

This article is organized as follows. In Sec.~\ref{sec:theory}, we  review the strong-field approximation and the CQSFA developed in \cite{Lai2015}, and give recent improvements in the latter. Subsequently, in Sec.~\ref{sec:qinterf}, we analyze near-threshold patterns in ATI, starting from those occurring in the SFA (Sec.~\ref{sec:interfsfa}). We then study the main types of orbits in the CQSFA (\ref{sec:CQSFAorbs}), provide analytic estimates for the PAD sidelobes (\ref{sec:sidelobes}) and inter-cycle interference (\ref{sec:analytic}), and link different types of orbits to several intra-cycle holographic structures (Sec.~\ref{sec:intraCQSFA}). Finally, in Sec.~\ref{sec:conclusions} we state our main conclusions.

\section{Background}
\label{sec:theory}
Our starting point is the time-dependent Schr\"odinger equation in atomic units
\begin{equation}
i\partial_t|\psi(t)\rangle=H(t)|\psi(t)\rangle \,,\label{eq:Schroedinger}
\end{equation}
which describes the evolution of an electron under the influence of the binding potential and the external field. The Hamiltonian $H(t)$ may be split into $H(t)=H_a+H_I(t)$, where 
\begin{equation}
H_a=\frac{\hat{\mathbf{p}}^{2}}{2}+V(\hat{\mathbf{r}})
\end{equation}
gives the field-free one-electron atomic Hamiltonian and $\hat{\mathbf{r}}$ and $\hat{\mathbf{p}}$ denote the position and momentum operators, respectively. We choose $V(\hat{\mathbf{r}})$ to be a Coulomb-type potential
\begin{equation}
V(\hat{\mathbf{r}})=-\frac{C}{\sqrt{\hat{\mathbf{r}}\cdot
		\hat{\mathbf{r}}}},\label{eq:potential}
\end{equation} 
where $0\leq C\leq 1$ is an effective coupling, and $H_I(t)$ gives the interaction with the external field. In the length and velocity gauge, $H_I(t)=-\hat{\mathbf{r}}\cdot \mathbf{E}(t)$ and $H_I(t)=\hat{\mathbf{p}}\cdot \mathbf{A}(t)+\mathbf{A}^2/2$, respectively, where $\mathbf{E}(t)=-d\mathbf{A}(t)/dt $
is the electric field of the external laser field and $\mathbf{A}(t)$ the corresponding vector potential. 
Eq.~(\ref{eq:Schroedinger}) can also be written in an integral form if we consider time evolution operators. This leads to the Dyson equation\begin{equation}
U(t,t_0)=U_a(t,t_0)-i\int^t_{t_0}U(t,t^{\prime})H_I(t^{\prime})U_a(t^{\prime},t_0)dt^{\prime}\,
,\label{eq:Dyson}
\end{equation}
where $U_a(t,t_0)=\exp[iH_a (t-t_0)]$ is the time-evolution operator associated with the field-free Hamiltonian, and the time evolution operator
\begin{equation}
U(t,t_0)=\mathcal{T}\exp \bigg [i \int^t_{t_0}H(t^{\prime})dt^{\prime} \bigg],
\end{equation}
where $\mathcal{T}$ denotes time-ordering, relates to the full Hamiltonian $H(t)$ evolving from an initial time $t_0$ to a final time $t$.

 In ionization, the quantity of interest is the transition amplitude $\left\langle\psi_{\textbf{p}}(t) |U(t,t_0) |\psi_{0} \right\rangle $ from a bound state $\left\vert \psi _{0}\right\rangle $ to a final continuum state $ |\psi_{\textbf{p}}(t)\rangle$ with momentum $\mathbf{p}$, which can be written in integral form using Eq.~(\ref{eq:Dyson}). This gives the formally exact ionization amplitude
\begin{equation}
M(\mathbf{p})=-i \lim_{t\rightarrow \infty} \int_{-\infty }^{t }d
t^{\prime}\left\langle \psi_{\textbf{p}}(t)
|U(t,t^{\prime})H_I(t^{\prime})| \psi _0(t^{\prime})\right\rangle \,
,\label{eq:transitionampl}
\end{equation}
 with $\left | \psi _0(t^{\prime})\right\rangle=\exp[iI_pt']\left\vert \psi _{0}\right\rangle $, where $I_p$ is the ionization potential.  Throughout, we will employ the length gauge, as it gives better results for ATI within the SFA \cite{Bauer2005PRA}. 
 
\subsection{Strong-field approximation}
\label{sec:SFA}
The strong-field approximation is obtained if the full time evolution operator is replaced by the Volkov time evolution operator $U^{(V)}(t,t^{\prime})$ in Eq.~(\ref{eq:transitionampl}). More detail is provided in \cite{Lohr1997,anatomy} and in the review article \cite{Popruzhenko2014JPB}. The main advantage is that this operator can be computed analytically. This however approximates the continuum by field-dressed plane waves, and thus eliminates the influence of the binding potential in the electron propagation.
 
Within the SFA, the transition amplitude for direct ATI from the initial bound state $|\psi_0\rangle$ to  a final Volkov state with drift momentum \textbf{p} is given by \cite{Becker_2002,Faria_2002, Keldysh_1980}
\begin{equation}
\label{Tamp}
M_{d}(\mathbf{p})=-i \int_{-\infty}^{\infty}dt' \langle \mathbf{p}+\mathbf{A}(t')|H_I(t')|\Psi_{0}\rangle e^{i S_{d}(\mathbf{p},t')},
\end{equation}
where
\begin{equation}  \label{Action}
S_d(\mathbf{p},t')= -\frac{1}{2} \int^{\infty}_{t'}[\mathbf{p}+\mathbf{A}(\tau)]^2 d\tau - I_pt'
\end{equation}
is the semiclassical action, which describes the propagation of an electron from the ionization time $t'$ to the end of the pulse, which is taken to be infinitely long. The electron's continuum state $|\mathbf{p}+\mathbf{A}(t')\rangle$ is a field-dressed plane wave with momentum $\mathbf{p}+\mathbf{A}(t')$, obtained by back propagating the final state $|\psi_{\textbf{p}}(t)\rangle$ from $t$ to $t'$ with $U^{(V)}(t',t)$. In Eqs (\ref{Tamp}) and (\ref{Action}), $I_p$ denotes the ionization potential. We use the length gauge Hamiltonian,  and employ the steepest descent method.  This means that we seek $t'$ for which Eq (\ref{Action}) is stationary, which gives the saddle point equation
\begin{equation} 
\frac{\partial S(t')}{\partial t'}=\frac{\lbrack \mathbf{p}+\mathbf{A}(t')]^{2}}{2}+ I_p=0.
\label{eq:Saddle}
\end{equation}

Physically, Eq (\ref{eq:Saddle}) ensures the conservation of energy upon tunnelling ionization at time $t'$ for the electron. Because tunnelling has no classical counterpart, this equation has only complex solutions. In terms of the solutions $t_s$ if Eq (\ref{eq:Saddle}), the transition amplitude (\ref{eq:Saddle}) can be approximated by
\begin{equation}
M_{d}(\mathbf{p})= \sum_s \mathcal{C}(t_s) e^{ S_{d}(\mathbf{p},t_s)},
\label{eq:Tamp2}
\end{equation}
where
\begin{equation}
\label{eq:Prefactor}
\mathcal{C}(t_s)=\sqrt{\frac{2 \pi i}{\partial^{2}S({\mathbf{p},t_s)} / \partial t^{2}_{s}}}\langle \mathbf{p}+\mathbf{A}(t_s)|H_I(t_s)|\Psi_{0}\rangle.
\end{equation}
The prefactor $\mathcal{C}(t_s)$ is expected to vary much more slowly than the action at each saddle for the saddle point approximation to hold \cite{Wahl_1964}. According to Eq.~(\ref{eq:Tamp2}), there are in principle many orbits along which the electron may be freed. This means that, for the same final momentum, the corresponding transition amplitudes will interfere.

\subsection{Coulomb quantum-orbit strong-field approximation}
\label{sec:CQSFA}
We will now insert the closure relation $\int d
\mathbf{\tilde{p}}_0 |\mathbf{\tilde{p}}_0\rangle\langle
\mathbf{\tilde{p}}_0 | =1$ in Eq.~(\ref{eq:transitionampl}). This gives 
\begin{eqnarray}\label{eq:Mpp}
M(\mathbf{p}_f)\hspace*{-0.1cm}&=&\hspace*{-0.1cm}-i \lim_{t\rightarrow \infty}\hspace*{-0.15cm}
\int_{-\infty }^{t }\hspace*{-0.2cm}d t'\hspace*{-0.2cm}
\int d \mathbf{\tilde{p}}_0 \left\langle  \mathbf{\tilde{p}}_f(t)
|U(t,t') |\mathbf{\tilde{p}}_0\right \rangle \nonumber \\
 && \times \left \langle
\mathbf{\tilde{p}}_0 | H_I(t')| \psi
_0(t')\right\rangle \, ,
\end{eqnarray}
where $|\mathbf{\tilde{p}}_f(t)\rangle=|\psi_{\mathbf{p}}(t)
\rangle$. The variables  $\mathbf{\tilde{p}}_0=\mathbf{p}_0+\mathbf{A}(t')$ and $\mathbf{\tilde{p}}_f(t)=\mathbf{p}_f+\mathbf{A}(t)$ 
give the initial and final velocity of the electron at the times $t'$ and $t$, respectively.  This specific formulation is very convenient, as $ \left\langle  \mathbf{\tilde{p}}_f(t)
|U(t,t') |\mathbf{\tilde{p}}_0\right \rangle $ can be computed using path-integral methods \cite{Kleinert2009,Milosevic2013JMP}.  One should note that the bound states of the system have been neglected in the above-stated closure relation, which, physically, corresponds to ignoring transitions between bound states.

The CQSFA transition amplitude then becomes
\begin{eqnarray}\label{1}
M(\mathbf{p}_f)&=&-i\lim_{t\rightarrow \infty
}\int_{-\infty}^{t}dt' \int d\mathbf{\tilde{p}}_0
\int_{\mathbf{\tilde{p}}_0}^{\mathbf{\tilde{p}}_f(t)} \mathcal {D}'
\mathbf{\tilde{p}}  \int
\frac{\mathcal {D}\mathbf{r}}{(2\pi)^3}  \nonumber \\
&& \times  e^{i S(\mathbf{\tilde{p}},\mathbf{r},t,t')}
\langle
\mathbf{\tilde{p}}_0 | H_I(t')| \psi _0  \rangle \, ,
\end{eqnarray}
where the action is given by
\begin{equation}\label{stilde}
S(\mathbf{\tilde{p}},\mathbf{r},t,t')=I_pt'-\int^{t}_{t'}[
\dot{\mathbf{p}}(\tau)\cdot \mathbf{r}(\tau)
+H(\mathbf{r}(\tau),\mathbf{p}(\tau),\tau]d\tau,
\end{equation}
and
\begin{equation}
H(\mathbf{r}(\tau),\mathbf{p}(\tau),\tau)=\frac{1}{2}\left[\mathbf{p}(\tau)+\mathbf{A}(\tau)\right]^2
+V(\mathbf{r}(\tau)).
\label{Hamiltonianpath}
\end{equation}
We compute the action along a two-pronged contour, and perform a series of approximations. The first part of the contour is taken to be parallel to the imaginary-time axis, going from $t'=t'_r+it'_i$ to $t'_r$. The second part of the contour is chosen to be along the real time axis, from $t'_r$ to $t$. Physically, the former and the latter arm of the contour describe tunnel ionization and the continuum propagation, respectively. The action then reads 
\begin{equation}
S(\mathbf{\tilde{p}},\mathbf{r},t,t')=S^{\mathrm{tun}}(\mathbf{\tilde{p}},\mathbf{r},t'_r,t')+S^{\mathrm{prop}}(\mathbf{\tilde{p}},\mathbf{r},t,t_r'),
\end{equation}
where $S^{\mathrm{tun}}(\mathbf{\tilde{p}},\mathbf{r},t'_r,t')$ and $S^{\mathrm{prop}}(\mathbf{\tilde{p}},\textbf{r},t,t'_r)$ give the action along the first and second part of the contour, respectively. This type of contour has been widely used in the literature \cite{Popruzhenko2008JMO,Yan2012,Torlina2012PRA,Torlina2013}. We assume the electron momentum to be approximately constant in the first arm of the contour. The explicit expressions for $S^{\mathrm{tun}}$ and $S^{\mathrm{prop}}$ are
\begin{eqnarray}
S^{\mathrm{tun}}(\mathbf{\tilde{p}},\mathbf{r},t'_r,t')&\hspace*{-0.2cm}=\hspace*{-0.2cm}&I_p(it'_i)-\frac{1}{2}\int_{t'}^{t'_r}\left[ \mathbf{p}_0+\mathbf{A} (\tau)\right]^2d\tau \notag \\ &&- \int_{t'}^{t'_r}V(\mathbf{r}_0(\tau))d\tau, \label{eq:stunn}
\end{eqnarray}where $\mathbf{r}_0$ is defined by 
\begin{equation}
\bm{r}_0(\tau)=\int_{t'}^{\tau}(\bm{p}_0+\bm{A}(\tau'))\text{d}\tau',
\label{eq:tunneltrajectory}
\end{equation}
and
\begin{eqnarray}
S^{\mathrm{prop}}(\mathbf{\tilde{p}},\mathbf{r},t,t'_r)&\hspace*{-0.2cm}=\hspace*{-0.2cm}&I_p(t_r)-\frac{1}{2}\int_{t'_r}^{t}\left[ \mathbf{p}(\tau)+\mathbf{A} (\tau)\right]^2d\tau \notag \\&&- \int_{t'_r}^{t}[\bm{\dot{p}}\cdot \bm{r} +V(\mathbf{r}(\tau))]d\tau, \label{eq:sprop}
\end{eqnarray}
respectively. The contour for $S^{\mathrm{tun}}(\mathbf{\tilde{p}},\mathbf{r},t'_r,t')$ inside the barrier will be computed from the origin until the tunnel exit, which is chosen as 
\begin{equation}\label{exit}
z_0=\Re[r_{0z}(t'_r)]
\end{equation}
as given in \cite{PPT1967}.

The above-stated equation will be solved by the stationary-phase method. In the CQSFA, we must seek solutions not only for the ionization time $t^{\prime}$ but also for the intermediate position and momentum $\mathbf{r}(\tau)$ and $\mathbf{p}(\tau)$ so that the action given by Eq.~(\ref{stilde}) is stationary.  This gives the equation
\begin{equation}
\frac{\left[\mathbf{p}(t')+\mathbf{A}(t')\right]^2}{2}+V(\mathbf{r}(t'))=-I_p,
\label{eq:tunncc}
\end{equation}
related to the energy conservation upon tunnel ionization, and
\begin{align}
\nabla_rS(\mathbf{\tilde{p}},\mathbf{r},t,t')&=0 \implies& \bm{\dot{p}}&=-\nabla_rV(\bm{r}(\tau)), \label{eq:q-spe} \\\nabla_pS(\mathbf{\tilde{p}},\mathbf{r},t,t')&=0 \implies&
	 \bm{\dot{r}}&= \bm{p}+A(\tau), \label{eq:p-spe}
\end{align}
which describe the dynamics of the electron in the continuum from $t'_r$ to $t$. Given $V(r)=- C/r$ we find
\begin{equation}
\mathbf{r} \cdot \dot{\mathbf{p}}=-\mathbf{r} \cdot \nabla_r V(r)=V(r). \label{eq:virial}
\end{equation}Hence, Eq.~(\ref{eq:virial}) can be substituted into Eq.~(\ref{eq:sprop}) to simplify it. This yields
\begin{eqnarray}
S^{\mathrm{prop}}(\mathbf{\tilde{p}},\mathbf{r},t,t'_r)&\hspace*{-0.2cm}=\hspace*{-0.2cm}&I_p(t_r)-\frac{1}{2}\int_{t'_r}^{t}\left[ \mathbf{p}(\tau)+\mathbf{A} (\tau)\right]^2d\tau \notag \\&&- 2\int_{t'_r}^{t}V(\mathbf{r}(\tau)))d\tau. \label{eq:sprop2}
\end{eqnarray}
This resembles the virial theorem, for which an analogous relationship between kinetic and potential energy can be derived. A similar result was obtained in \cite{Shvetsov-ShilovskiPRA2016}.

In Eq.~(\ref{eq:stunn}), we have approximated the momentum to be fixed. We can thus neglect the binding potential in Eq.~(\ref{eq:tunncc}) which gives the ionization time. This leads to 
\begin{equation}
\frac{1}{2}\left[ \mathbf{p}_0+\mathbf{A}(t')\right]^2+I_p=0.
\label{eq:tp-spe}
\end{equation}
The potential is however included in the equations of motion (\ref{eq:q-spe}) and (\ref{eq:p-spe}) and in the action (\ref{eq:sprop}), which are solved for a specific final momentum $\mathbf{p}_f$ and $t \rightarrow \infty$. 
The initial momenta are computed by solving  Eqs.~(\ref{eq:q-spe}) and (\ref{eq:p-spe}), with the tunnel exit as an initial position and the final momenta as a final ``limit'' condition. In order to implement the limit on the momenta we  solve the problem iteratively, starting from the SFA and increasing the influence of the Coulomb potential. This method does not explicitly parametrize the initial momenta in terms of the final, but enables each orbit's initial momentum to be calculated for any given final momentum. This makes it much easier to see the momentum distributions for each orbit, which gives a unique insight into the dynamics.

After a series of manipulations, the Coulomb corrected transition amplitude becomes
{\compeqn
	\begin{equation}
	\label{eq:MpPathSaddle}
	M(\mathbf{p}_f)\propto-i \lim_{t\rightarrow \infty } \sum_{s}\bigg\{\det \bigg[  \frac{\partial\mathbf{p}_s(t)}{\partial \mathbf{r}_s(t_s)} \bigg] \bigg\}^{-1/2} \hspace*{-0.6cm}
	\mathcal{C}(t_s) e^{i
		S(\mathbf{\tilde{p}}_s,\textbf{r}_s,t,t_s))} ,
	\end{equation}}where $t_s$, $\bm{p}_s$ and $\bm{r}_s$ are determined by Eqs.~(\ref{eq:q-spe})-(\ref{eq:tp-spe}) and $\mathcal{C}(t_s)$ is given by Eq.~(\ref{eq:Prefactor}).  In practice, we employ  the stability factor $\partial
\mathbf{p}_s(t)/\partial \mathbf{p}_s(t_s)$, which is obtained using a Legendre transformation. The action will remain the same as long as the
electron starts from the origin. Eq.~(\ref{eq:MpPathSaddle}) is normalized so that the SFA transition amplitude is obtained in the limit of vanishing binding potential. Throughout, we consider the electron to be initially bound in a $1s$ state. For details we refer to \cite{Lai2015}.  

\section{Quantum-interference effects}
\label{sec:qinterf}
All the different orbits that reach the detector with the same final momenta will contribute to the interference patterns. These orbits and the corresponding patterns will be the main topic of this section, in the SFA and CQSFA. For simplicity, in the results that follow, we will consider a linearly polarized monochromatic field 
\begin{equation}
\mathbf{E}(t)=E_0 \sin(\omega t)\hat{e}_{z}.
\label{eq:efield}
\end{equation}
This corresponds to the vector potential
\begin{equation}
\mathbf{A}(t)=2 \sqrt{U_p} \cos(\omega t)\hat{e}_{z},
\label{eq:afield}
\end{equation}
where $\hat{e}_{z}$ gives the unit vector in the direction of the driving-field polarization and $U_p$ is the ponderomotive energy.  In our studies, we will focus on the action as it plays the most important role in determining phase differences between quantum orbits. The prefactors vary much more slowly, and will only play a secondary role.  
Under this approximation, the probability distribution considering $N_{c}$ cycles of the driving field and a number $n_{e}$ of relevant events per cycle is given by 
\begin{equation}
\Omega(\mathbf{p}_f)=\left|\sum_{e=1}^{n_{e}}
\sum_{c=1}^{N_c}\exp[iS_{ec}]\right|^2,
\label{eq:generalinterf}
\end{equation}
where $S_{ec}$ is the action associated to the $e$-th event in the $c$-th cycle and $\mathbf{p}_f$ the momentum at the detector. Here the single sum over $s$ in Eq.~(\ref{eq:MpPathSaddle}) has been replaced by a double sum  in the indices $e$ and $c$. 

For a monochromatic field we find that the difference
\begin{equation}
S_{ec'}-S_{ec}=\frac{2\pi i (c'-c)}{\omega}\left(\ip+\up+\frac{1}{2}\bm{p}_f^2\right)
\label{eq:intercycle}
\end{equation}
between the actions related to the same type of orbit but a different cycle  is independent of the orbit. This renders Eq.~(\ref{eq:generalinterf}) factorizable and given by
\begin{equation}
\Omega(\mathbf{p}_f)= 
\Omega_{n_e}(\mathbf{p}_f)\Omega_{N_c}(\mathbf{p}_f), 
\label{eq: generalizedinterf}
\end{equation}
where $\Omega_{n_e}(\mathbf{p}_f)$ is the probability associated with intra-cycle interference and 
\begin{equation}
	\Omega_{N_c}(\mathbf{p}_f)=\frac{\cos\left[\frac{2\pi i N_c}{\omega}\left(\ip+\up+\frac{1}{2}\bm{p}_f^2\right)\right]-1}{\cos\left[\frac{2\pi i }{\omega}\left(\ip+\up+\frac{1}{2}\bm{p}_f^2\right)\right]-1} \label{eq:nintercycle}
\end{equation}
is the probability related to inter-cycle interference. Details about Eqs.(\ref{eq:generalinterf})-(\ref{eq:nintercycle}) are provided in Appendix \ref{app:generalized}. In the limit of infinitely long pulses, Eq.~(\ref{eq:nintercycle}) describes a Dirac delta comb, whose peaks are unequally spaced, and remains the same for the SFA and CQSFA. This condition agrees with the expression in \cite{ArboNuclInstr}.

The number $n_e$ of relevant orbits per cycle will depend on the approach, and will not exceed three in this work. Hence, we can write an expression for intra-cycle interference that is general enough to encapsulate all the effects discussed. Explicitly,
{\compeqn
\begin{equation}
	\Omega_{n_e}(\bm{p}_f)=e^{-2\Im[S_{1c}]}\left|1
	+e^{- \Delta S^{\Im}_{12}} e^{i \Delta S^{\Re}_{1 2}}
	+e^{- \Delta S^{\Im}_{13}} e^{i \Delta S^{\Re}_{1 3}}
\right|^2.
\end{equation}}Here $\Delta S^{\Re}_{1j}=\Re[S_{jc}-S_{1c}]$ and $\Delta S^{\Im}_{1j}=\Im[S_{jc}-S_{1c}]$ with $j=2,3$. This is valid for all cycles $c$. The term $e^{-2\Im[S_{1c}]}$ shapes the momentum distribution, and gives rise to the side-lobes identified in ATI photoelectron momentum distributions \cite{Huismanset2010Science}. The real parts of $\Delta S_{ij}$ lead to the interference fringes seen in ATI, while $\mathrm{Im}[\Delta S_{ij}]$  switch interference on or off. If  $\mathrm{Im}[\Delta S_{ij}]$ is small,  then interference is on, whereas for large $\mathrm{Im}[\Delta S_{ij}]$ interference is off and one of the orbits prevails.

We will next write the action for the monochromatic fields (\ref{eq:efield}) and (\ref{eq:afield}).
The tunnelling and propagation parts of the action in the CQSFA given in Eq.~(\ref{eq:stunn}) and (\ref{eq:sprop2}), respectively, can be rewritten as
\begin{align}
&\hspace*{-1mm}S^{\mathrm{tun}}(\bm{\tilde{p}},\bm{r},t_r',t')= i\left(\ip+\frac{1}{2}\bm{p}_0^2 + \up \right) t'_i  -\int^{t'_r}_{t'} V(\bm{r}_0(\tau))\mathrm{d}\tau \nonumber\\& \hspace{1.8cm}
+ 
\frac{2\sqrt{\up}p_{0 z}}{\omega} \left[ \sin(\omega t')-\sin(\omega t'_r)\right]\nonumber \\ & \hspace{1.8cm}
+\frac{\up}{2\omega}\left[\sin(2\omega t') -\sin(2\omega t'_r) \right]
 \label{eq:stunn_ex}
 \end{align}
and
\begin{align}
&\hspace{-3mm}S^{\mathrm{prop}}(\bm{\tilde{p}},\bm{r},t,t'_r)=
\left(\ip + \frac{1}{2}\bm{p}_f^2 + \up \right) t'_r 
+ \frac{2\sqrt{\up}p_{fz}}{\omega}\sin(\omega t'_r) \nonumber \\&\hspace{13mm}
+\frac{\up}{2\omega}\sin(2\omega t'_r) 
-\frac{1}{2}\int_{t'_r}^{t}\pmb{\mathscr{P}}(\tau)\cdot (\pmb{\mathscr{P}}(\tau)+2\bm{p}_f)\mathrm{d}\tau \nonumber \\&\hspace{13mm}
-2\sqrt{\up}\int_{t'_r}^{t}\mathscr{P}_{z}(\tau)\cos(\omega \tau)\mathrm{d}\tau 
-2 \int^{t}_{t'_r} V(\bm{r}(\tau))\mathrm{d}\tau, \label{eq:sprop_ex}
\end{align}
where $p_{jz}$, with $j=0,f$, correspond to the electron momentum components parallel to the laser-field polarization and  $\bm{p}(\tau)=\pmb{\mathscr{P}}(\tau)+\bm{p}_f$.  This has been chosen so that all the integrands go to zero for large $\tau$. Eq.~(\ref{eq:stunn_ex}) and (\ref{eq:sprop_ex}) can be combined to give an explicit form of the action,
\begin{align}
&S(\bm{\tilde{p}},\bm{r},t,t')
=\left(\ip + \up \right) t'
+\frac{1}{2}\bm{p}^2_f t'_{r}+\frac{i}{2}\bm{p}^2_{0}  t'_{i}
+\frac{\up}{2\omega}\sin(2\omega t') \nonumber \\& \hspace{-2.5mm}
+ \frac{2\sqrt{\up}}{\omega}\left[ p_{ 0z}\sin(\omega t')-(p_{0z}-p_{fz})\sin(\omega t'_{r}) \right] \hspace*{-1mm}
-\hspace*{-1mm}\int^{t'_{r}}_{t'} \hspace*{-1mm}V(\bm{r}_{0}(\tau))\mathrm{d}\tau \nonumber\\& \hspace{-2.5mm}
 -\frac{1}{2}\int_{t'_{r}}^{t}\pmb{\mathscr{P}}(\tau)\cdot (\pmb{\mathscr{P}}(\tau)+2\bm{p}_f+2\bm{A}(\tau))\mathrm{d}\tau
-2 \int^{t}_{t'_{r}} \hspace*{-1mm}V(\bm{r}(\tau))\mathrm{d}\tau. \label{eq:s_ex}
\end{align}
This equation can be considered general in that we will recover the SFA if the Coulomb coupling is reduced  to zero, i.e., in the limit $C\rightarrow 0$. Then $V(\bm{r})\rightarrow 0$, $\bm{p}_f\rightarrow \bm{p}_0\rightarrow\bm{p}$ and $\pmb{\mathscr{P}}\rightarrow 0$, which leaves us with the SFA action given in Eq.~(\ref{eq:ActionSFA}).

\subsection{Strong-field approximation}
\label{sec:interfsfa}
For the SFA, Eq.~(\ref{eq:s_ex}) gives an explicit form of the action, if the above limits are taken,
\begin{eqnarray}
\label{eq:ActionSFA}
S_d(\mathbf{p},t')&=&\left(\frac{p^2_{z}+p^2_{x}}{2}+I_p+U_p\right)t' \\ 
&&+\frac{2 p_{z}\sqrt{U_p}}{\omega}\sin[\omega t']+\frac{U_p}{2 \omega}\sin[2\omega t'],\notag
\end{eqnarray}
where $p_{z}$ and $p_{x}$ correspond to the momentum components parallel and perpendicular to the laser-field polarization, which remain constant throughout (i.e., $\mathbf{p}_0=\mathbf{p}_f=\mathbf{p}$). The saddle-point equation (\ref{eq:Saddle}) can be solved analytically for the ionization time $t_{ec}$ related to an event $e$ occurring in a cycle $c$. This gives
\begin{align}
t_{ec} = \frac{2\pi n}{\omega}\pm\frac{1}{\omega} \arccos\left(\frac{-p_{z}\mp i\sqrt{2 I_\mathrm{p}+p^2_{x}}}{2\sqrt{U_\mathrm{p}}}\right),
\label{eq:timesSFA}
\end{align}
where $n$ is any integer.  Convergent solutions require that $\mathrm{Im}[t_{ec}]>0$.  This parametrization has been used in \cite{Kopold2000}
\subsubsection{Interference condition}
Within the SFA, the dominant types of interference are determined by two ionization events, occurring at the times $t_{ec}$ and $t_{e'c'}$. The number of events in a cycle is restricted to $e=1,2$ so that they relate to orbit 1 and 2, respectively. The dominant interference patterns occur for the condition
\begin{equation}
\mathrm{Re}[t_{ec}]\pm \mathrm{Re}[t_{e'c'}]=2\pi/\omega.
\label{eq:interfcondition}
\end{equation}
If $e=e'$, $c \neq c'$ and the negative sign is chosen, Eq.~(\ref{eq:interfcondition}) describes the dominant inter-cycle interference contributions. One should note, however, that there are also secondary events  separated by more than a cycle, which can be obtained by considering $2n_c\pi/\omega$, $n_c>1$, on the right-hand side of Eq.~(\ref{eq:interfcondition}). For intra-cycle interference, one must take $e \neq e'$, $c=c'$ and the positive sign in Eq.~(\ref{eq:interfcondition}).

Furthermore, 
\begin{equation}
\mathrm{Im}[t_{ec}]=\mathrm{Im}[t_{e'c'}],
\label{eq:interfimaginary}
\end{equation}
which, physically, reflects the fact that, in the SFA, the potential barrier is determined solely by the driving field. 

The quantity of interest is 
\begin{equation}
	\Omega(\mathbf{p})=|e^{iS_{ec}}(1+e^{i\Delta S})|^2, 
\end{equation}
where $S_{ec}=S(t_{ec})$ is the SFA action ~(\ref{eq:ActionSFA}) associated with each of the interfering events and $\Delta S=S_{e'c'}-S_{ec}$ is the corresponding phase difference. 
The real part of $\Delta S$ gives the interference fringes, while its imaginary part determines the contrast of the patterns.  Eq.~(\ref{eq:interfimaginary}) guarantees sharp fringes as $\mathrm{Im}[\Delta S]=0 $.

For intercycle interference, condition (\ref{eq:interfcondition}) gives
\begin{equation}
	\Delta S^{\mathrm{(SFA)}} _{\mathrm{inter}}= \left(\frac{p^2_{z}+p^2_{x}}{2}+I_p+U_p\right) \frac{2 n_c\pi}{\omega}.
	\label{eq:interSFA}
\end{equation}
Interference extrema requires that 
\begin{equation}
	p^2_{z}+p^2_{x}=\frac{n}{n_c}\omega-2U_p-2I_p, 
	\label{eq:SFAcondition2}
\end{equation}
where even and odd $n$ give maxima and minima, respectively. This condition describes a circle centered at $(p_{z},p_{x})=(0,0)$, and it is exact within the SFA framework. For the radius in Eq.~(\ref{eq:SFAcondition2}) to be real,  $n \omega \geq 2 n_c (U_p+I_p)$. The dominant processes correspond to the shortest time difference, i.e.,  $n_c=1$. For $n_c > 1$, the fringes will be much finer and start at a higher value of $n$. For a coherent superposition of all inter-cycle processes, the interference condition follows Eq.~(\ref{eq:nintercycle}). 

For intracycle interference, we use the specific solutions
\begin{align}
t_{1c}&= \frac{1}{\omega} \arccos\left(\frac{-p_{z}- i\sqrt{2 I_\mathrm{p}+p^2_{x}}}{2\sqrt{U_\mathrm{p}}}\right) \label{eq:t1s}\\
t_{2c} &= \frac{2\pi}{\omega}-\frac{1}{\omega} \arccos\left(\frac{-p_{z}+ i\sqrt{2 I_\mathrm{p}+p^2_{x}}}{2\sqrt{U_\mathrm{p}}},\right) \label{eq:t2s}
\end{align}
which are both in the upper complex half plane.   For $p_{z}>0$, Eqs.~(\ref{eq:t1s}) and (\ref{eq:t1s}) are related to orbits 1 and 2. For orbit 1, the electron is released in the direction of the detector and for orbit 2 it is released in the opposite direction, which is subsequently changed by the field. For $p_{z}<0$, the situation is reversed and the solutions are shifted in half a cycle \cite{Yan2010,Lai2015}.
\begin{figure}[t] 
	\includegraphics[width=\columnwidth]{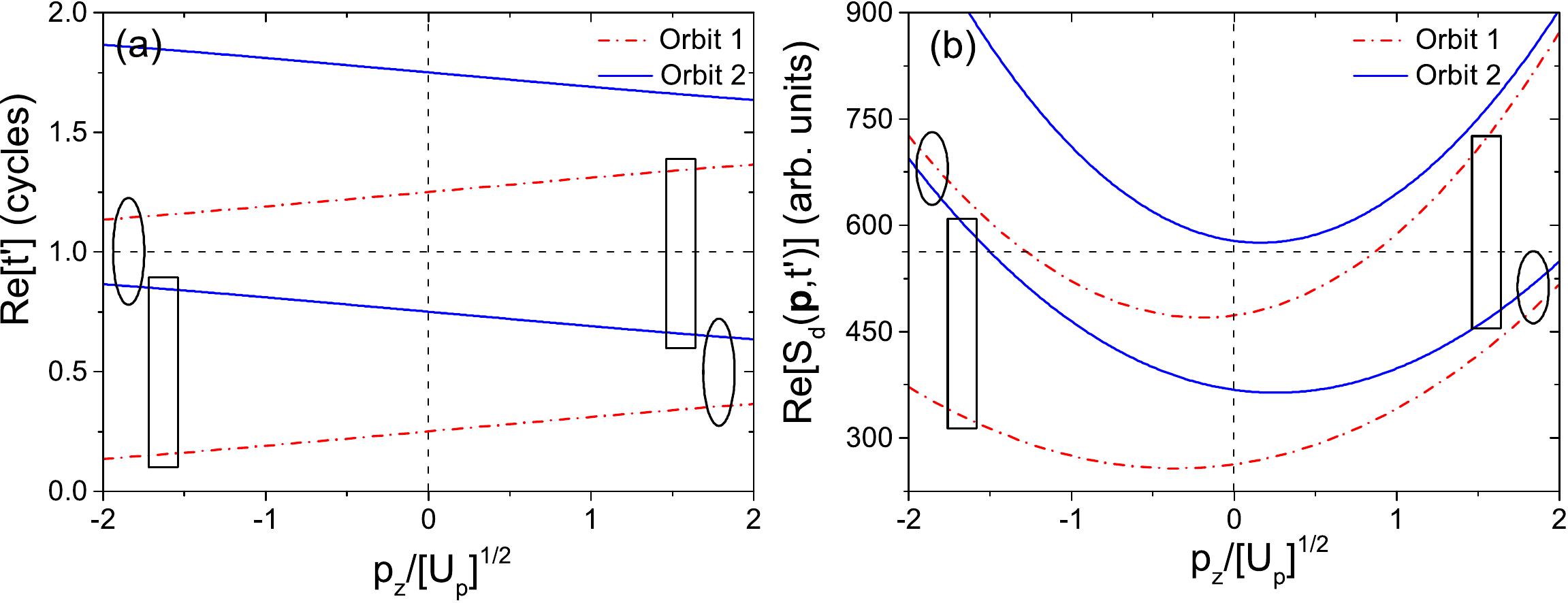}
	\caption{Real part of the ionization times $t_{ec}$ and actions $S_{ec}$ as functions of the electron momentum component $p_{z}$ parallel to the driving-field polarization [panels (a) and (b), respectively].   In panel (a), from top to bottom, we see the real parts of the solutions $t_{11}$, $t_{21}$, $t_{12}$ and $t_{22}$, respectively, while panel (b) displays the real parts of the actions  $S_{11}$, $S_{21}$, $S_{12}$ and $S_{22}$. The circles (squares) indicate the interfering parts of the orbits for which $\Delta t$ is less than (greater than) half a cycle, which lead to type A (type B) intracycle interference. We have taken the perpendicular momentum component $p_x$ to be vanishing and renormalized $p_{z}$ in terms of $\sqrt{U_p}$. }\label{fig:MomReTimes}
\end{figure}

Applying conditions (\ref{eq:interfcondition}) and (\ref{eq:interfimaginary}) parametrizing $\omega t_{1c}= \arccos \xi$ according to Eq.~(\ref{eq:t1s}) gives 
 \begin{align}
 \Delta S^{\mathrm{(SFA)}}_{\mathrm{intra}} &=\left(I_\mathrm{p}+U_\mathrm{p}+\frac{1}{2}p_{x}^2+\frac{1}{2}p_{z}^2\right)\left(\frac{2\pi}{\omega}- \frac{2 \mathrm{Re}[\arccos(\xi)]}{\omega}\right) \notag \\
 &+\frac{4 p_{z}\sqrt{U_\mathrm{p}}}{\omega} \mathrm{Re}\left[\sqrt{1-\xi^2}\right]
 +\frac{2 U_\mathrm{p}}{\omega} \mathrm{Re}\left[\xi \sqrt{1-\xi^2}\right].
 \label{eq:deltaSintra}
 \end{align}
 Interference maxima and minima require $\Delta S^{\mathrm{(SFA)}}_{\mathrm{intra}}= n \pi$, for $n$ even and odd, respectively.  Eq.~(\ref{eq:deltaSintra}) can be used to describe two types of interference patterns. If $\mathrm{Re}[t_{2c}-t_{1c}]$ is smaller (greater) than half a cycle, we will refer to type A (type B) intra-cycle interference, respectively.
 A schematic representation of orbits 1 and 2, together with the corresponding actions, is provided in Fig.~\ref{fig:MomReTimes}. Type A  interference has been extensively studied in the literature, while type B interference has been overlooked.

\begin{figure*}[t]
	\includegraphics[width=13cm]{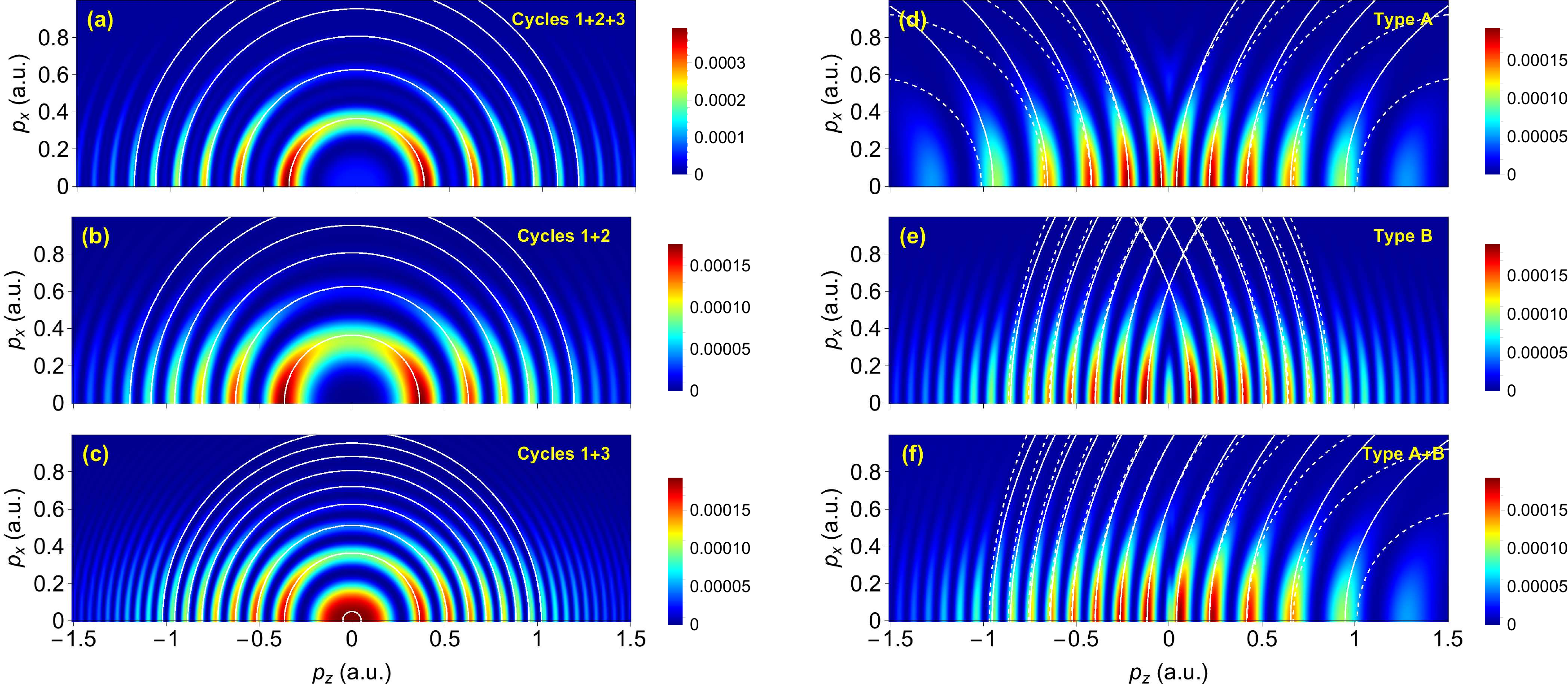}
	\caption{Electron momentum distributions computed with the SFA for Hydrogen $I_p=0.5$ a.u. in a driving field of intensity $I=2 \times 10^{14} \mathrm{W/cm}^2$ and frequency $\omega=0.057$ a.u.. In panels (a) to (c), we display inter-cycle interference patterns obtained using orbit 1. In Panel (a), we consider the ionization times $t_{11}$, $t_{12}$ and $t_{13}$. In panel (b) we take only events within the first two cycles, i.e., the times $t_{11}$ and $t_{12}$, while panel (c) was computed using only $t_{11}$ and $t_{13}$.  The solid white lines superimposed to the fringes in panel (a) give the analytic condition in Eq.~(\ref{eq:nintercycle}) with $N_c=3$, and those in panels (b) and (c) follow  Eq.~(\ref{eq:SFAcondition2})  with  $n_c=1$ and $n_c=2$, respectively. In  panels (d) to (f), we present intracycle interference patterns computed using the times $t_{11}$ and $t_{21}$. Panels (d) and (e) exhibit type A and B intracycle interference, for which  $\Delta t$ is less than or greater than half a cycle, respectively. Panel (f) was computed following the solutions $t_{11}$ and $t_{21}$ from negative to positive parallel momenta without imposing any restriction upon the time difference. This gives type A intracycle interference for $p_{z} >0$ and type B intracycle interference $p_{z} < 0$. The solid and dashed white lines superimposed to the fringes give the exact and approximate SFA conditions for intra-cycle interference [Eqs.~(\ref{eq:deltaSintra}) and (\ref{eq:SFAcondition1})], respectively. } \label{fig:interfSFA1}
\end{figure*}
It is helpful to derive approximate intra-cycle  conditions by expanding $\mathrm{Re}[t_{1c}]$ and $\mathrm{Re}[t_{2c}]$ around two consecutive field extrema. To zeroth order,  $\mathrm{Re}[t_{1c}]=\pi/(2\omega)+ n\pi/\omega$ and $\mathrm{Re}[t_{2c}]=(2n+3)\pi/(2\omega)$. In this case, the electron reaches the continuum with vanishing momentum, i.e., $p_{z}=p_{x}=0$. This gives $\xi_0=i\sqrt{I_p/(2U_p)}$, which, if inserted in Eq.~(\ref{eq:deltaSintra}) leads to
\begin{equation}
 \left(p_{z}-\frac{4\sqrt{U_p}}{\pi}\right)^2\hspace*{-0.2cm}+p^2_{x}=2n\omega-2U_p-2I_p+\hspace*{-0.08cm}\left(\frac{4\sqrt{U_p}}{\pi}\right)^2,
 \label{eq:SFAcondition1}
\end{equation}
for $\Delta S^{\mathrm{(SFA)}}_{\mathrm{intra}}=n\pi$, which is the equation of a circle centered at $(p_{z},p_{x})=(4\sqrt{U_p}/\pi,0)$.
 
For both types of intra-cycle interference, the approximate condition (\ref{eq:SFAcondition1}) works well around the origin, but worsens for increasing momentum components, while Eq. (\ref{eq:deltaSintra}) is exact within the SFA framework. We access the negative momentum regions by considering $p_{z} \rightarrow -p_{z}$ in both equations, which corresponds to a shift of half a cycle in the solutions $t_{ec}$. This gives another circle centered at $(p_{z},p_{x})=(-4\sqrt{U_p}/\pi,0)$. Using Eq.~(\ref{eq:SFAcondition1}), one may determine a range for the interference order $n$, within which type A interference may occur. The condition that the radius in Eq. (\ref{eq:SFAcondition1}) must be positive gives the lower bound $n\omega > U_p+I_p-8U_p/\pi^2$ for $n$. Furthermore, if the electron leaves at a field crest, one may set $p_{z}=p_{x}=0$ in Eq.~(\ref{eq:SFAcondition1}). This yields the upper bound  $	n \leq I_p + U_p$. One should note that, for type B interference, the latter expression constitutes an approximate lower bound for $n$.
 
 \subsubsection{Interference patterns}

 Fig.~\ref{fig:interfSFA1} displays photoelectron angular distributions (PADs)                                                                                                                                                                                                                                                                                                                                                                                                                                                                                                                                                                                                                                                                                                                                                                                                                                                                                                                                                                                                                                                                                                                                                                                                                                                                                                                                                                                                                                                                                                                                                                                                                                                                                                                                                                                                                                                                                                                                                                                                                                                                                                                                                          constructed so that specific types of inter- and intracycle interference are isolated. In Fig.~\ref{fig:interfSFA1}(a), we show ring-shaped patterns from a coherent superposition of type 1 orbits within three field cycles. The rings are modulated and follow Eq.~(\ref{eq:nintercycle}) with $N_c=3$, which suggests a coherent superposition of two types of rings. This is confirmed by considering only the first and the second cycle, for which $n_c=1$ in Eq.~(\ref{eq:SFAcondition2}) [Fig.~\ref{fig:interfSFA1}(b)], or the first and the third cycle, for which $n_c=2$ [Fig.~\ref{fig:interfSFA1}(c)]. For larger $n_c$, the fringes start at higher momentum and are finer, as expected from Eq.~(\ref{eq:SFAcondition2}). 
  
The remaining panels display intra-cycle interference. In Fig.~\ref{fig:interfSFA1}(d), we plot type A intra-cycle interference, using the pairs of orbits marked by the circles in Fig.~\ref{fig:MomReTimes}. Note that, for different signs of $p_{z}$, the chosen solutions have been shifted by half a cycle. This leads to symmetric patterns with regard to $p_{z}\rightarrow -p_{z}$. As the momentum $p_{z}$ increases in absolute value, the corresponding ionization times move from two consecutive field extrema $(\Delta t=\pi/\omega)$ towards the same field crossing $(\Delta t=0)$. For that reason, the phase difference $\Delta S$ decreases [see black circles in Fig.~\ref{fig:MomReTimes}(b)], which leads to broader fringes in the angle-resolved spectra as $|p_{z}|$ increases. Fig.~\ref{fig:interfSFA1}(e) depicts type B intra-cycle interference, using the orbits indicated by the rectangles in Fig.~\ref{fig:MomReTimes}. In this case, the real parts of the ionization times move from two consecutive field extrema  towards different field crossings as $|p_{z}|$ increases. Thus, $\Delta S$ increases and the fringes  become finer. If one follows a specific pair of solutions from negative to positive $p_{z}$ relaxing the above constraints upon $\Delta t$, this results in the momentum distribution presented in Fig.~\ref{fig:interfSFA1}(f). The smooth decrease in $\Delta S$ leads to a gradual transition from finer to thicker interference fringes. 

\begin{figure}[t]
	\includegraphics[width=\columnwidth]{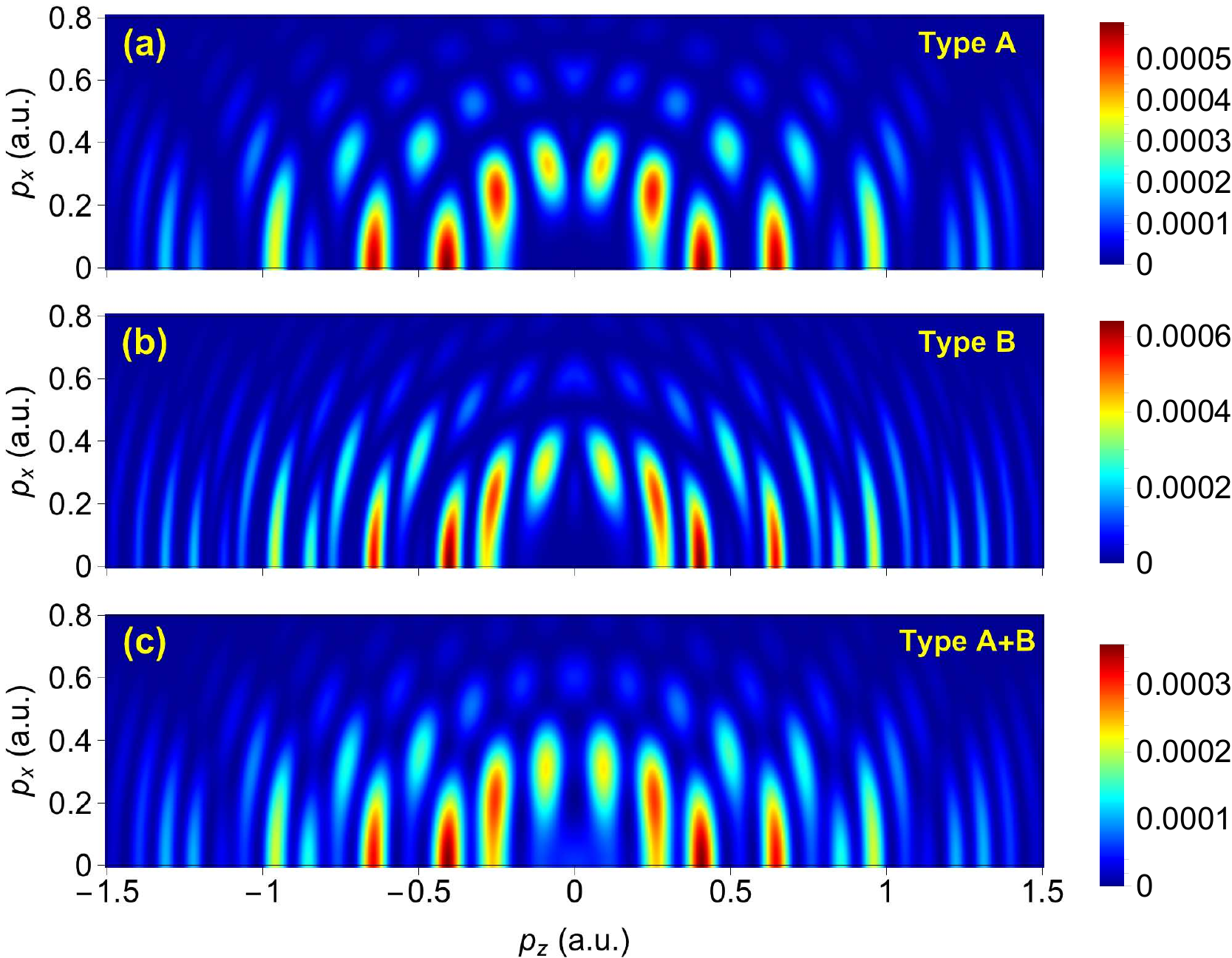}
	\caption{Momentum distributions computed with the SFA using the same field and atomic parameters as in Fig.~\ref{fig:interfSFA1}, for two cycles of the fundamental driving field. Panel (a) was constructed using orbits 1 and 2 for $p_{z}>0$ and symmetrization with regard to the origin, which allows for type A and B intra-cycle interference occurring twice and once, respectively. This implies choosing the times $t_{11}$, $t_{21}$ and $t_{12}$ for $p_{z}>0$ and symmetrizing with regard to $p_{z}=0$. Physically, this symmetrization entails shifting the unit cells in half a cycle for $p_{z}<0$. In panel (b), we allow for types A and B intra-cycle interference to occur once and twice, respectively. This can be achieved by shifting the times used in panel (a) by half a cycle, i.e.,  employing $t_{21}$, $t_{12}$ and $t_{22}$ for $p_{z}>0$ and symmetrizing with regard to $p_{z}=0$. In panel (c) we consider two consecutive orbits 1 and only one orbit 2 over 2.5 cycles, which gives an equal number of times for type A or B interference occurring. In all three plots intercycle interference rings appear. This is because more than one cycle is considered, which introduces interference between consecutive orbits 1 and 2. } \label{fig:interfSFA3}
\end{figure}
A real pulse is however composed of at least a few cycles, so that all types of interference will be present. Fig.~\ref{fig:interfSFA3} provides three examples of angle-resolved photoelectron distributions computed over two cycles of the fundamental driving field. Overall, we see the intercycle interference rings in the momentum maps, but the shapes of the intra-cycle fringes are determined by the dominant events. In Figs.~\ref{fig:interfSFA3}(a) and (b), we construct the patterns such that type A or type B intracycle interference prevails, respectively. For that reason, in Fig.~\ref{fig:interfSFA3}(a) the outward curves at either side of $p_{z}=0$ dominate, while in Fig.~\ref{fig:interfSFA3}(b) the intra-cycle fringes become finer and turn inward near $p_{z}=0$.  In Fig.~\ref{fig:interfSFA3}(c) both types of interference are included on equal footing. This leads to straight vertical lines at either side of $p_{z}=0$ as the curvatures of the types A and B interference outweigh each other. A long enough pulse leads to approximately symmetric distributions. However, exact symmetry only occurs for an infinitely long, monochromatic wave.
\subsection{Coulomb-corrected approach}
\label{sec:interfCQSFA}
\subsubsection{Quantum Orbits}
\label{sec:CQSFAorbs}
In the following we will have a closer look at the orbits that exist in the CQSFA. In Coulomb corrected models of ATI there are four types of orbits for any given momenta. Their standard characterization is based on the tunnel exit $z_0$ and the initial transverse momenta $p_{0x}$ with regard to the final parallel and transverse momenta $p_{fz}$ and $p_{fx}$, respectively \cite{Yan2010}. For  orbit 1,  $z_0$ and the electron's final momentum $p_{fz}$ point in the same direction, i.e.,  $z_0p_{fz} > 0$, and its initial and final transverse momenta have the same  sign, i.e., $p_{ 0x} p_{ fx} > 0$. Orbits 2 and 3 have their tunnel exit on the opposite side, so that $z_0 p_{fx} < 0$. Orbit 2  has its initial transverse momentum in the same direction as the final momentum ($p_{0x} p_{fx} > 0$), while for orbit 3 these momentum components point in opposite directions ($p_{0x} p_{fx} < 0$). Finally, orbit 4 has its tunnel exit on the same side as $p_{fz} $, but the initial and final transverse momenta are in opposite directions, i.e., $p_{0x} p_{fx} < 0$. The transition amplitude related to orbit 4 is small, hence we will not consider it any further \cite{Lai2015}. This characterization differs from that employed in Sec. \ref{sec:SFA}, as the solutions $t_{ec}$ associated with each orbit are not kept continuous for all momenta. Keeping  $t_{ec}$  continuous would change the behavior of the orbits according to this classification, which we would like to avoid. 

One of the main differences between the SFA and CQSFA is that momenta do not remain constant in the latter. Hence, one can no longer assume that two orbits with the same initial momenta will interfere, as they may reach the detector with different final momenta.
\begin{figure*}
	\includegraphics[width=1.\linewidth]{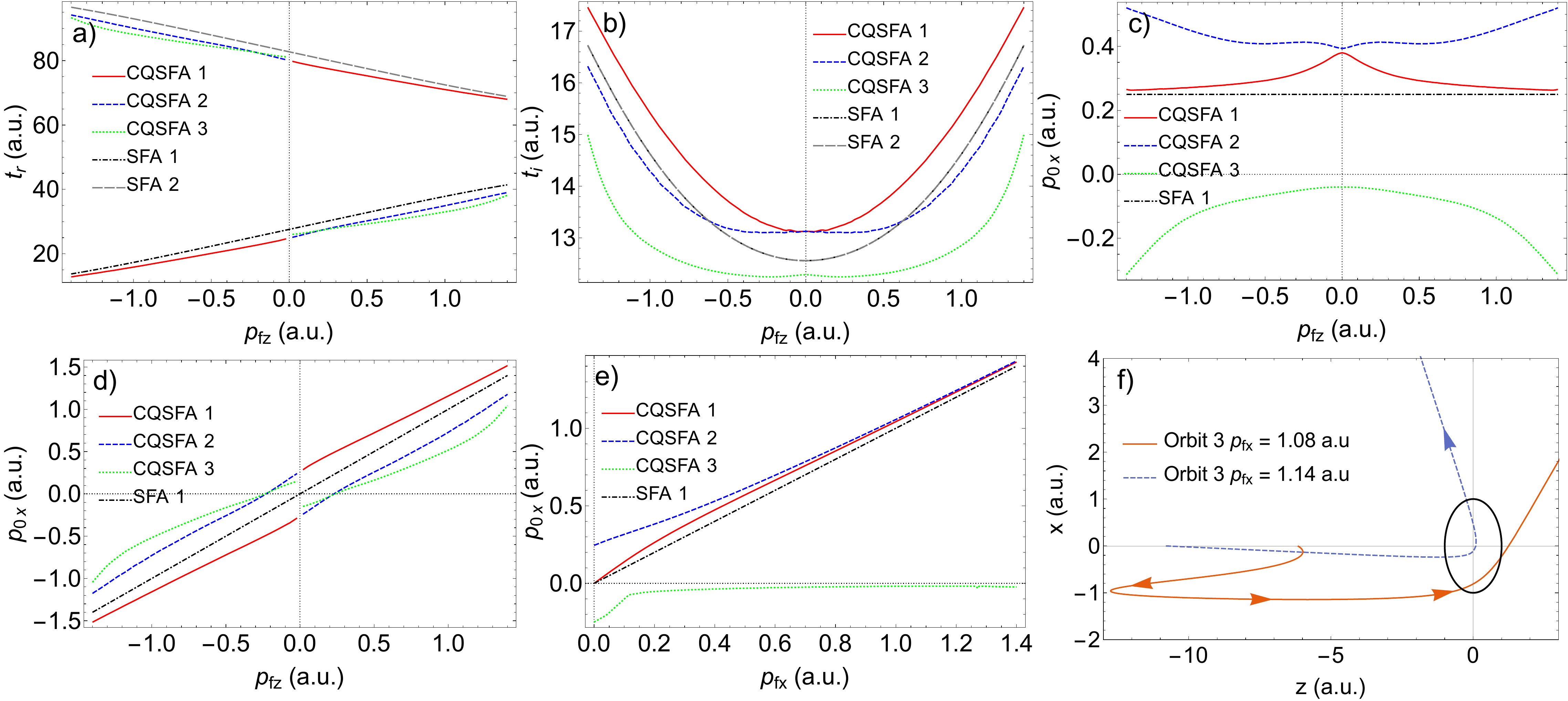}
	\caption{In panels (a) and (b), we plot the real and imaginary part of the ionization times obtained for the CQSFA orbits 1, 2 and 3, as functions of the final momentum component $p_{fz}$ parallel to the laser-field polarization, compared with their SFA counterpart (black and gray lines in the figure)  		Panels (c) and (d) show the initial perpendicular and parallel momentum components $p_{0x}$ and $p_{0z}$ for the CQSFA orbits 1 to 3, respectively, as functions of the final parallel momentum $p_{fz}$. In panels (a) to (d), the final perpendicular momentum was chosen as  $p_{fx}= 0.25$ a.u.. Panel (e) presents the initial perpendicular momentum $p_{0x}$ as a function of the final perpendicular momenta  $p^{(i)}_{fx}$, $i=1,2,3$, for a fixed value of $p_{fz}= 0.25$ a.u.. For reference, from panels (c) to (e) the SFA solution is provided as the black dotted-dashed line.  Panel (f) shows orbit 3 for two values of the initial perpendicular momentum. For a final momentum  $(p_{fx},p_{fz}) = (1.082 \hspace*{0.1cm}\mathrm{a.u.},0.668\hspace*{0.1cm} \mathrm{a.u.})$ and an initial momentum of $(p_{0x},p_{0z}) = (-0.043\hspace*{0.1cm}\mathrm{a.u.},0.563\hspace*{0.1cm}\mathrm{a.u.})$ (solid orange line), the electron deflected by the potential, while for  $(p_{fx},p_{fz}) =		(1.144\hspace*{0.1cm}\mathrm{a.u.},0.672\hspace*{0.1cm}\mathrm{a.u.})$ and an initial momentum of $(p_{0x},p_{0z}) =(-0.041\hspace*{0.1cm}\mathrm{a.u.},2.713\hspace*{0.1cm}\mathrm{a.u.})$ (dashed blue line), the electron undergoes a hard collision with the core. The black circle in the figure marks the region for which the collision occurs.  The field and atomic parameters are the same as in Figs.~\ref{fig:interfSFA1} and \ref{fig:interfSFA3}.  }
	\label{Fig:Orbits}
\end{figure*}
The ionization times, like in the SFA, can be explicitly parametrized in terms of the initial momenta. This leads to an orbit-dependent version of Eq.~(\ref{eq:timesSFA}), with the SFA momentum $\mathbf{p}$  replaced by the initial CQSFA momentum $\mathbf{p}_0$. For $p_{fz}>0$, the times $t_{1c}$ associated with orbit 1 are given by  Eq.~(\ref{eq:t1s}), with $\mathbf{p}$ replaced by $\mathbf{p}^{(1)}_0$, while those related to orbits 2 and 3 are given by Eq.~(\ref{eq:t2s}), with  $\mathbf{p}$ replaced by $\mathbf{p}^{(2)}_0$ or $\mathbf{p}^{(3)}_0$, respectively.  Differences between the times $t_{2c}$ and $t_{3c}$ for orbit 2 and 3 come from the fact that they have different initial momenta. For $p_{fz}<0$, the situation reverses, i.e., $t_{1c}$ is given by Eq.~(\ref{eq:t2s}) and the remaining times by Eq.~(\ref{eq:t1s}).

In Fig.~\ref{Fig:Orbits}(a), we display the real  parts of the ionization times as functions of the electron's final momentum $p_{fz}$ parallel to the laser-field polarization, which are associated to the classical trajectories of an electron in the field. We can see from Fig.~\ref{Fig:Orbits}(a) that the real part of the time of ionization for the CQSFA is quite similar to the SFA but is shifted down. Physically, this can be understood as follows: For orbit 1, the electron is decelerated by the Coulomb potential, so that it will need a higher momentum $\mathbf{p}_0$ to escape and reach the detector with a specific momentum $\mathbf{p}_f$. This means that the driving field must compensate the above-mentioned deceleration and that the electron's release time $t_{1c}$ must move away from the field extremum towards the crossing. In contrast, for orbits 2 and 3 the binding potential accelerates the electron and it must acquire less energy from the field to achieve a final momentum $\mathbf{p}_f$. Thus, the electron is released with a lower momentum and its release times must approach the previous field extremum. As $|p_{fz}|$ increases, all three times tend to their SFA counterparts, but reach this limit in different ways.  The time $t_{1c}$ tends monotonically towards the SFA value, while the ionization times $t_{2c}$ and $t_{3c}$ first deviate from their SFA counterparts. This is because an electron along orbit 1 may escape with vanishing transverse momentum $p_{0x}=0$, while for orbits 2 and 3 this would either trap the electron or lead to a hard rescattering with the core in case the $p_{0z}$ is low.

In Fig.~\ref{Fig:Orbits}(b), we show the imaginary parts $\mathrm{Im}[t_{ec}]$, with $e=1,2,3$, of these solutions.
An overall feature is that they are no longer identical, so that Eq.~(\ref{eq:interfimaginary}) breaks down for intra-cycle events. This is expected, as $\mathrm{Im}[t_{ec}]$ is roughly related to the width of the effective potential barrier through which the electron tunnels \cite{Faria2004}. The Coulomb potential will make this barrier different for orbits 1, 2 and 3, while in the SFA it is determined solely by the field.  Qualitatively, $\mathrm{Im}[t_{1c}]$ behaves in the same way for the SFA and CQSFA, with a clear minimum at $p_{fz}=0$. This is not surprising, as the topology of orbit 1 is similar in both cases. In contrast, for orbit 2, $\mathrm{Im}[t_{2c}]$ exhibits a maximum at $p_{fz}=0$ and two symmetric minima at non-vanishing momenta. This effect is quite robust, and contributes to the appearance of side lobes in the PADs. For orbit 3, $\mathrm{Im}[t_{3c}]$ is much flatter and smaller than for the other two orbits, which  indicates a high escape probability over a large momentum range. This is consistent with the electron being accelerated for a longer time, in comparison to orbit 2. Similarly to what occurs for orbit 2, $\mathrm{Im}[t_{3c}]$ exhibits a local maximum for $p_{fz}=0$ and two symmetric minima at $p_{fz} \neq 0$. There is however a sharp increase in $\mathrm{Im}[t]$ for higher parallel momenta, as hard collisions with the core start to take place [see Fig.~\ref{Fig:Orbits}(f)]. This regime is outside the scope of this work, and will not be addressed here. 

In Fig.~\ref{Fig:Orbits}(c), we plot the initial parallel momenta as functions of the final perpendicular momentum. For orbit 1, if the electron escapes along the polarization axis, it will need an initial momentum corresponding to the classical escape velocity $\sqrt{2C/|z_0|}$, determined by setting $|V(z_0)|=v_{0z}^2/2$. For non-vanishing transverse momentum, analytical estimates for the escape velocity are non-trivial. Still, the figure clearly shows a monotonic decrease in $p^{(1)}_{0z}$.
Orbits 2 and 3, on the other hand, need a much lower momentum  to escape and reach the detector along the polarization axis. Thus, $p_{0z}$ eventually increases with final transverse momentum. 

Similar features are observed in Fig.~\ref{Fig:Orbits}(d), where $p^{(e)}_{0z}$, $e=1,2,3$ are displayed as functions of $p_{fz}$.  Importantly, orbit 1 never crosses the $p_{fz}$ axis.  This is because the electron starts with the atomic potential directly behind it. Hence, it must have a large enough initial parallel velocity to be able to escape. Furthermore, for  orbits 2 and 3, the SFA solution $p_{0z}=p_{fz}$ is approached from below, while for orbit 1 it is approached from above.  This is a consequence of the electron being accelerated by the potential along the two former orbits, and decelerated along the latter. The acceleration is more significant for orbit 3, in agreement with the previous plots.  The critical behavior of this orbit is also shown  in Fig.~\ref{Fig:Orbits}(e)  in which  $p_{0x}$ is plotted as function of its final value $p_{fx}$. For orbits 1 and 2, the SFA value is reached when the momentum increases, but this does not happen for orbit 3. 
 
 \subsubsection{Single-cycle distributions and side lobes}
 \label{sec:sidelobes}
 \begin{figure}
 	\hspace*{-0.8cm}	\includegraphics[width=10cm]{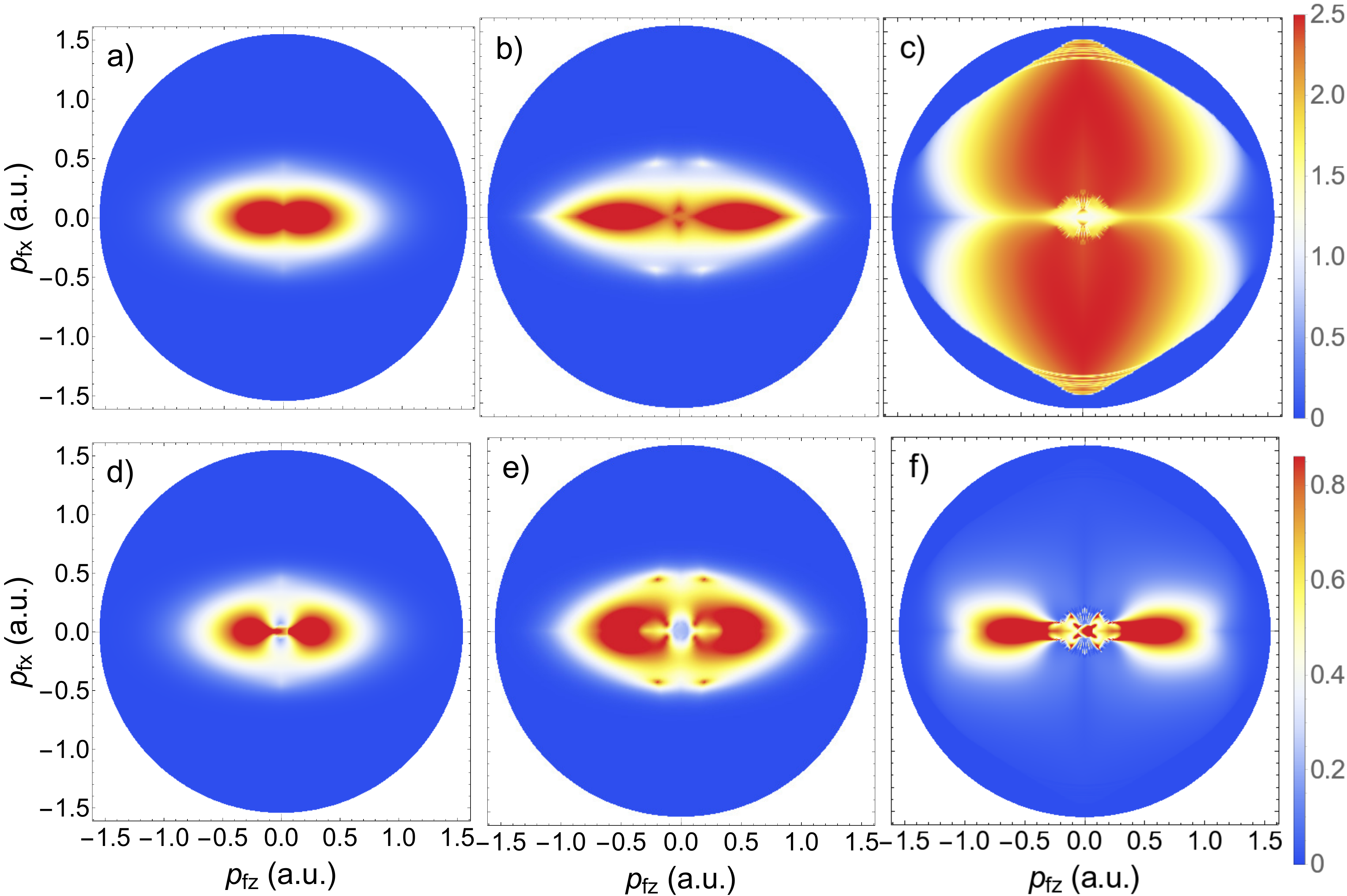}
 	\caption{Single-orbit angle-resolved probability distributions plotted in arbitrary units and computed for the same field and atomic parameters as in the previous figures. The  left, middle and right columns correspond to orbit 1, 2 and 3, respectively. The upper and  panels have been computed using solely the actions, while in the lower panels we have included the prefactors. The upper panels have been multiplied by $10^3$ in order to facilitate a comparison with the lower ones.}
 	\label{fig:SingleOrbit}
 \end{figure}
  In Fig.~\ref{fig:SingleOrbit}, we plot the PADs computed using single orbits. In the upper panels we consider only the influence of the action, while in the lower panels we include the whole prefacor, given by the stability factor mulitplied by $\mathcal{C}(t_s)$ in Eq.~(\ref{eq:MpPathSaddle}).  Overall, we see the presence of sidelobes for the contributions of orbits 1 and 2. They mainly stem from the imaginary part of the action [Figs.~\ref{fig:SingleOrbit}(a) and (b)] but are enhanced by the prefactors [Figs.~\ref{fig:SingleOrbit}(d) and (e)].  Furthermore, in Fig.~\ref{fig:SingleOrbit}(c), one can see that the contributions of orbit 3 decay more slowly than those of the two other orbits. Around 1.2 a.u. there is a sharp decay in probability, as above a certain energy an electron leaving along orbit 3 starts to backscatter. This prominence is however obfuscated by the influence of the prefactor, which causes a huge suppression of the probability density away from the $p_{fz}$ axis [Figs.~\ref{fig:SingleOrbit}(f)]. This led to us neglecting this orbit in previous work  \cite{Lai2017}. 
  
\begin{figure}
 	\includegraphics[width=0.48\textwidth]{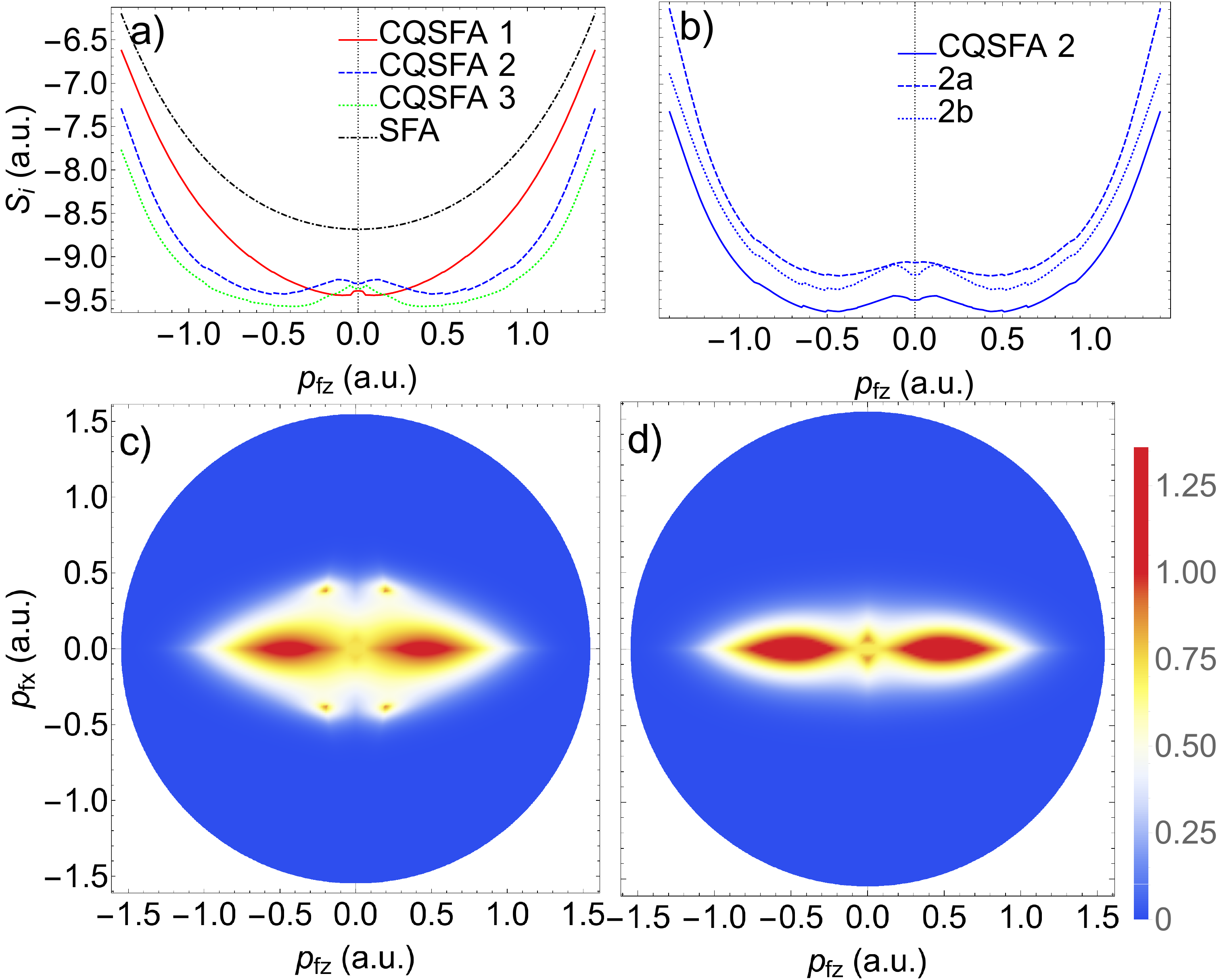}
 	\caption{Panel (a) shows the imaginary parts of the actions $S_i$ (i=1,2,3) associated to the orbits 1, 2 and 3 of the CQSFA, as functions of the final parallel momentum, computed directly from Eq.~(\ref{eq:ImCQSFAAction}) for perpendicular momentum $p_{fx}=0.05$ a.u.. For comparison, the SFA counterpart has been included. Panel (b) displays the approximate expressions obtained  for orbit 2, as functions of the final parallel momentum, for the same perpendicular momentum as panel (a). The dotted line, labelled 2b, corresponds to the single-orbit action without the integral over the binding potential, and the dashed lines, labelled 2a, include this integral in the long-wavelength approximation. The solid line gives the numerical expression for Eq.~(\ref{eq:ImCQSFAAction}). Panels (c) and (d) illustrate the PADs computed for orbit 2 without the prefactors, with and without the integral over  $V(\bm{r}_{0}(\tau))$ in the long-wavelength approximation, respectively. The atomic and field parameters are the same as in the previous figures. }
 	\label{fig:ImAction}
 \end{figure}

In the CQSFA, the imaginary part of the action reads 
 {\vcompeqn
 	\begin{align}
 S^{\text{Im}}(t',\bm{p},\bm{r})&= \left(I_p+U_\mathrm{p}+\frac{1}{2}\bm{p}_0^2\right)t'_{i}
 	+\frac{2 p_{0z} \sqrt{U_\mathrm{p}} \cos (\omega t'_{r} )\sinh (\omega t'_{i} )}{\omega }\nonumber\\
 	&	+\frac{U_\mathrm{p}\cos (2  \omega t'_{r} )\sinh (2 \omega t'_{i}) }{2 \omega }
 	-\int^{t'_{r}}_{t'} \Im[V(\bm{r}_{0}(\tau))]\mathrm{d}\tau.
 	\label{eq:ImCQSFAAction}
 	\end{align}
 }
 Eq.~(\ref{eq:ImCQSFAAction}) is plotted in Fig.~\ref{fig:ImAction}(a), for orbits 1, 2 and 3. In general, its behavior mirrors that observed for the imaginary parts of the ionization times.  This includes it being much smaller and flatter for orbit 3 and the local minima outside the origin for orbit 2.  
 
 The mirroring behavior can be seen from Eq.~(\ref{eq:ImCQSFAAction}) if one applies the low-frequency approximation \cite{Yan2012}. This gives $\sinh (\omega t'_i)\simeq \omega t'_i$ and $\sinh (2\omega t'_i)\simeq 2\omega t'_i$, which is the dominant term. Within the same approximation, the integral over $V(\bm{r}_{0}(\tau))$ leads to an algebraic term, which may be viewed as a modified prefactor and whose influence is secondary as far as the sidelobes are concerned. It does however play an important role in the overall shape of the distributions. The explicit derivation of this term is presented in Appendix \ref{app:integral1}. 
 
 In Fig.~\ref{fig:ImAction}(b), we plot the action $S^{\text{Im}}(t_{2c},\bm{p},\bm{r})$ associated with orbit 2, including or not the integral over $V(\bm{r}_{0}(\tau))$ in the low-frequency approximation. In all cases, the two minima are present. Examples of single-orbit PADs computed analytically are provided in Figs.~\ref{fig:ImAction}(c) and (d). Both figures show clear side lobes and resemble the single-orbit distribution in Fig.~\ref{fig:SingleOrbit}(b), which has been computed numerically. However, inclusion of the integral over the binding potential in the low-frequency approximation renders the numerical and analytical single-orbit distributions strikingly similar. This similarity includes the broader shape and secondary peaks. 

\subsubsection{Intercycle interference}
\label{sec:analytic}
In the following, we will show that the expression for intercycle interference remains the same for the CQSFA, provided the field is monochromatic.  Using Eq.~(\ref{eq:interfimaginary}) and the field periodicity, the CQSFA action difference may be written as
\begin{equation}
\Delta S_{\mathrm{inter}}=\Delta S^{\mathrm{(SFA)}}_{\mathrm{inter}}+\Delta S_{cc'}, 
\end{equation}
where the first term refers to Eq.~(\ref{eq:interSFA}) with $\mathbf{p}$ replaced by $\mathbf{p}_f$, and $\Delta S_{cc'}$ are Coulomb corrections related to an event of the type $e$ occurring in cycles $c$ and $c'$, so that the ionization times satisfy $t'_{c'}=t'_{c}+2\pi n_c/\omega$. The indices $e$ are dropped as the condition refers to the same type of orbit. This action difference reads 
\begin{equation}
\Delta S_{cc'}=\Delta S_{V_T}+\Delta S_{V_C}+\Delta S_{p},
\end{equation} 
where $\Delta S_{V_T}$ and $\Delta S_{V_C}$ are the phase differences caused by the potential during tunnelling and continuum propagation, respectively, and $\Delta S_{p}$ is related to the change in momentum during the electron propagation. Explicitly, 
\begin{equation}
\Delta S_{V_T} = \int^{t'_{c'r}}_{t'_{c'}} V(\bm{r}_{c'0}(\tau))\mathrm{d}\tau-\int^{t'_{cr}}_{t'_{c}} V(\bm{r}_{c0}(\tau))\mathrm{d}\tau,
\label{eq:DeltaSVt}
\end{equation}
where the subscripts $r$ indicate the real parts of $t'_c$ and $t'_{c'}$ and $\mathbf{r}_{0c}(\tau)$ is given by Eq.~(\ref{eq:tunneltrajectory}) with the lower bound replaced by $t'_c$.
For a monochromatic field,  $\bm{r}_{c'0}(\tau)=\bm{r}_{c0}(\tau-\frac{2\pi n}{\omega})$. This may be used to show that the first and the second integrals cancel out, so that Eq.~(\ref{eq:DeltaSVt})  vanishes. 

The action difference 
\begin{equation}
\Delta S_{V_C}=-2\int^{t}_{t'_{c'r}} V(\bm{r}_{c'}(\tau))\mathrm{d}\tau
+2\int^{t}_{t'_{cr}} V(\bm{r}_c(\tau))\mathrm{d}\tau 
\end{equation}
is handled in a similar way, using  $\bm{r}_{c'}(\tau)=\bm{r}_{c}(\tau-\frac{2\pi n}{\omega})$. This gives 
\begin{equation}
\Delta S_{V_C} =2\int^{t}_{t-2\pi/\omega} \hspace*{-0.3cm} V(\bm{r}_c(\tau))\mathrm{d}\tau,
\end{equation}
which vanishes in the limit of $t \rightarrow \infty$. The same procedure, together with the mapping $\bm{p}_{c'}(\tau)=\bm{p}_{c}(\tau-\frac{2\pi n}{\omega})$, can also be used to show that 
\begin{eqnarray}
\Delta S_{p}&=&-\frac{1}{2}\int_{t'_{c'r}}^{t}\hspace*{-0.2cm} \pmb{\mathscr{P}}_{c'}(\tau)\cdot (\pmb{\mathscr{P}}_{c'}(\tau)+2\bm{p}_f+2\bm{A}(\tau))\mathrm{d}\tau \notag\\&&+\frac{1}{2}\int_{t'_{cr}}^{t}\hspace*{-0.2cm} \pmb{\mathscr{P}}_{c}(\tau)\cdot (\pmb{\mathscr{P}}_{c}(\tau)+2\bm{p}_f+2\bm{A}(\tau))\mathrm{d}\tau
\end{eqnarray}
vanishes in this limit. Hence, the Coulomb potential has no effect on the ATI rings.
\subsubsection{Intracycle interference}
\label{sec:intraCQSFA}
\begin{figure}
	\includegraphics[width=0.5\textwidth]{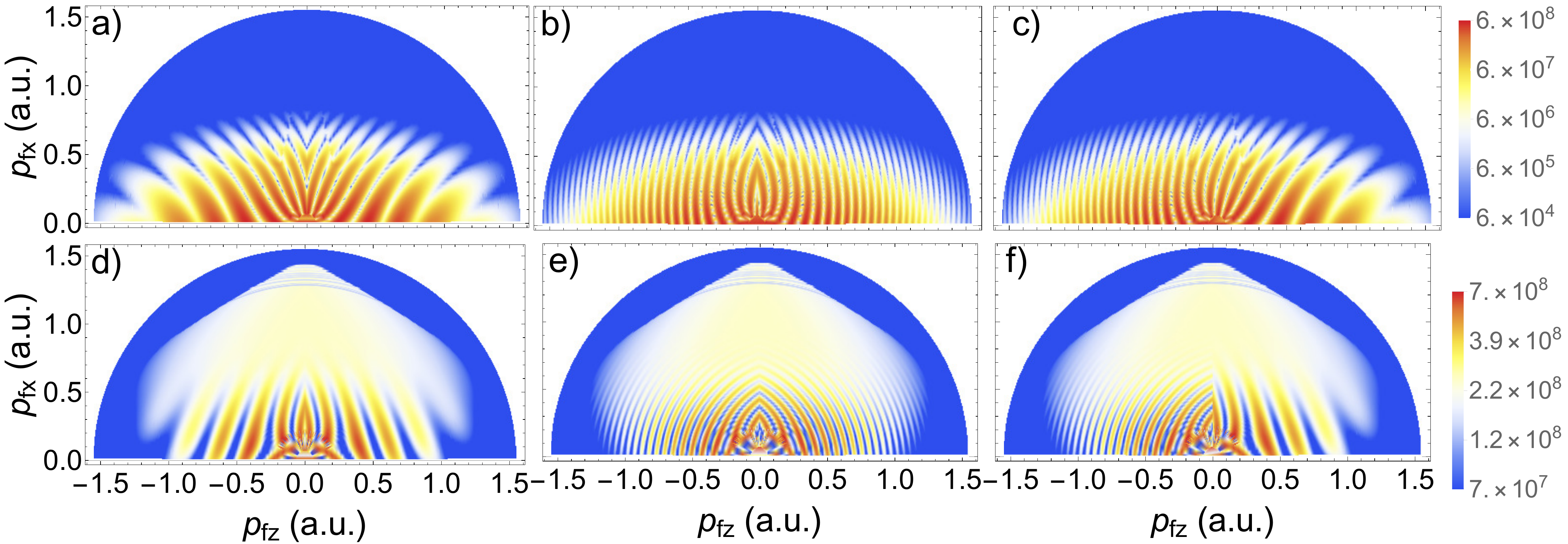}
	\caption{Photoelectron angular distributions computed in the CQSFA for times within a single cycle of the laser field and the same parameters as in the previous figures neglecting the prefactors. The upper and bottom row includes orbits 1 and 2, and orbits 1 and 3 as interfering trajectories, respectively. Panels (a) and (d) show type A intra-cycle interference,  panels (b),  and (e) present type B intra-cycle interference and panels (c), and (f) exhibit both types of interference, obtained in a similar way as in Fig.~\ref{fig:interfSFA1} by not imposing temporal constraints upon the interfering solutions. The panels have been plotted in a logarithmic scale.  }
	\label{fig:A_B_ABCQSFA}
\end{figure}
\begin{figure}
	\includegraphics[width=0.5\textwidth]{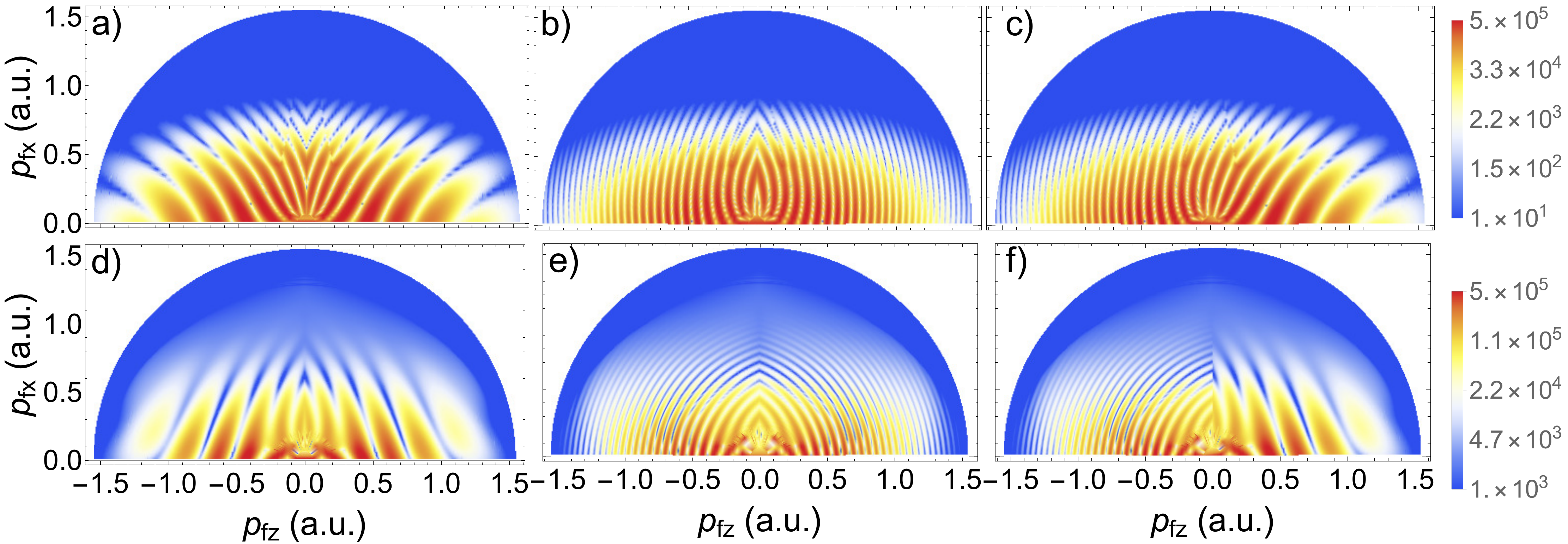}
	\caption{Photoelectron angular distributions computed in the CQSFA for times within a single cycle of the laser field and the same parameters as in the previous figures including the prefactors.  The upper and bottom row includes orbits 1 and 2, and orbits 1 and 3 as interfering trajectories, respectively. Panels (a) and (d) show type A intra-cycle interference,  panels (b),  and (e) present type B intra-cycle interference and panels (c), and (f) exhibit both types of interference, obtained in a similar way as in Fig.~\ref{fig:interfSFA1} by not imposing temporal constraints upon the interfering solutions. The panels have been plotted in a logarithmic scale.  }
	\label{fig:A_B_ABCQSFApref}
\end{figure}
\begin{figure}
	\includegraphics[width=0.5\textwidth]{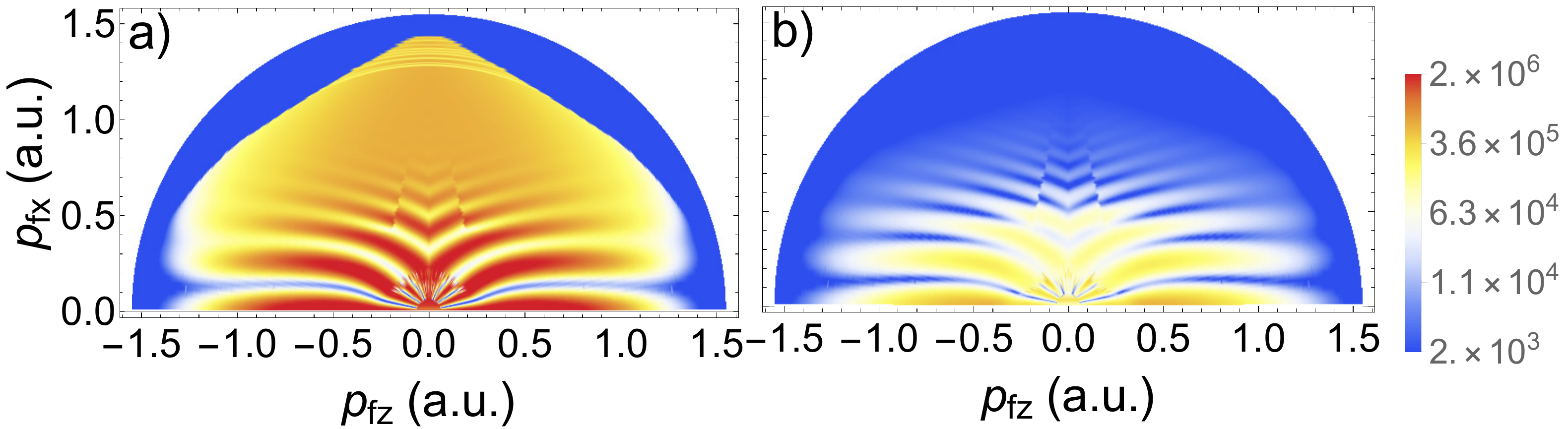}
	\caption{Photoelectron angular distributions computed in the CQSFA using orbits 2 and 3 for the same parameters as in Figs.~\ref{fig:A_B_ABCQSFA} and \ref{fig:A_B_ABCQSFApref} without and with prefactor [panels (a) and (b), respectively]. The panels have been plotted in a logarithmic scale.  }
	\label{fig:intra23}
\end{figure}

Figs.~\ref{fig:A_B_ABCQSFA} and \ref{fig:A_B_ABCQSFApref} exemplify the types of interference that occur in the CQSFA, with and without the full prefactor, respectively.  The left, middle and right panels in both figures refer to type A, type B and type A and B intracycle interference, respectively, computed in a similar fashion as for the SFA [right column in Fig.~\ref{fig:interfSFA1}]. The patterns obtained are more complex than those in the SFA, as there are three interfering types of orbits. Furthermore, since the imaginary parts $\textrm{Im}[t_{ec}]$ differ for each type of orbit, the fringes may become blurred in specific momentum regions. 

If only orbits 1 and 2 are taken [upper panels of Figs.~\ref{fig:A_B_ABCQSFA} and \ref{fig:A_B_ABCQSFApref}], the fringes are sharp and the fringe spacing is similar to that observed in the SFA. This is expected, as  $\textrm{Im}[t_{1c}]$ and $\textrm{Im}[t_{2c}]$ are comparable and $\textrm{Re}[t_{1c}]$ and $\textrm{Re}[t_{2c}]$ follow the SFA solutions closely. The shapes of the distributions, however, are different. Specifically, for type A intra-cycle interference, instead of the  nearly vertical fringes in Fig.~\ref{fig:interfSFA1}(b), we see a fan-shaped structure spreading from the origin $(p_{fz},p_{fx})=(0,0)$ [Figs.~\ref{fig:A_B_ABCQSFA}(a) and \ref{fig:A_B_ABCQSFApref}(a)]. This structure is well known, both theoretically and experimentally.
 Type B interference, shown in Figs.~\ref{fig:A_B_ABCQSFA}(b) and \ref{fig:A_B_ABCQSFApref}(b),  exhibits sharp, nearly vertical fringes, which resemble those observed for the SFA but also become distorted for low momentum regions. If both types of interference are considered, once more the fringes become increasingly thicker as the momenta move from the negative to the positive $p_{fz}$ region.
The presence of the prefactor enhances the side lobes, but does not change these features.

The interference between orbits 1 and 3, shown in the lower panels of Figs.~\ref{fig:A_B_ABCQSFA} and \ref{fig:A_B_ABCQSFApref}, behaves in a different way. First, the shapes of the fringes do not resemble the finger-shaped strucures or those from the SFA and the side lobes are absent. Second, if the prefactors are absent [Fig.~\ref{fig:A_B_ABCQSFA}], they are only sharp near the $p_{fz}$ axis and up to $p_{fx}\simeq 0.5$. For higher perpendicular momenta, the fringes are blurred and the PADs acquire the shape of the single-orbit distribution in Fig.~\ref{fig:SingleOrbit}(c). This is due to the  high probability of an electron leaving along orbit 3.  In Fig.~\ref{fig:A_B_ABCQSFApref}, however, one can see that the prefactor outweighs this high probability and suppresses the contribution of orbit 3 away from the $p_{fz}$ axis. If the intra-cycle interference between orbits 2 and 3 is considered (Fig.~\ref{fig:intra23}), we observe a set of prominent, almost horizontal fringes diverging from a spider-like structure near the origin. A similar structure has been observed in \cite{LiPRL2014} using the QMTC method. The prefacor restricts the relevance of this structure to a relatively narrow momentum range close to the $p_{fz}$ axis. One should note that, since these specific orbits leave in the same half cycle, the classification in A and B type interference is not applicable.

\begin{figure}
	\includegraphics[width=0.5\textwidth]{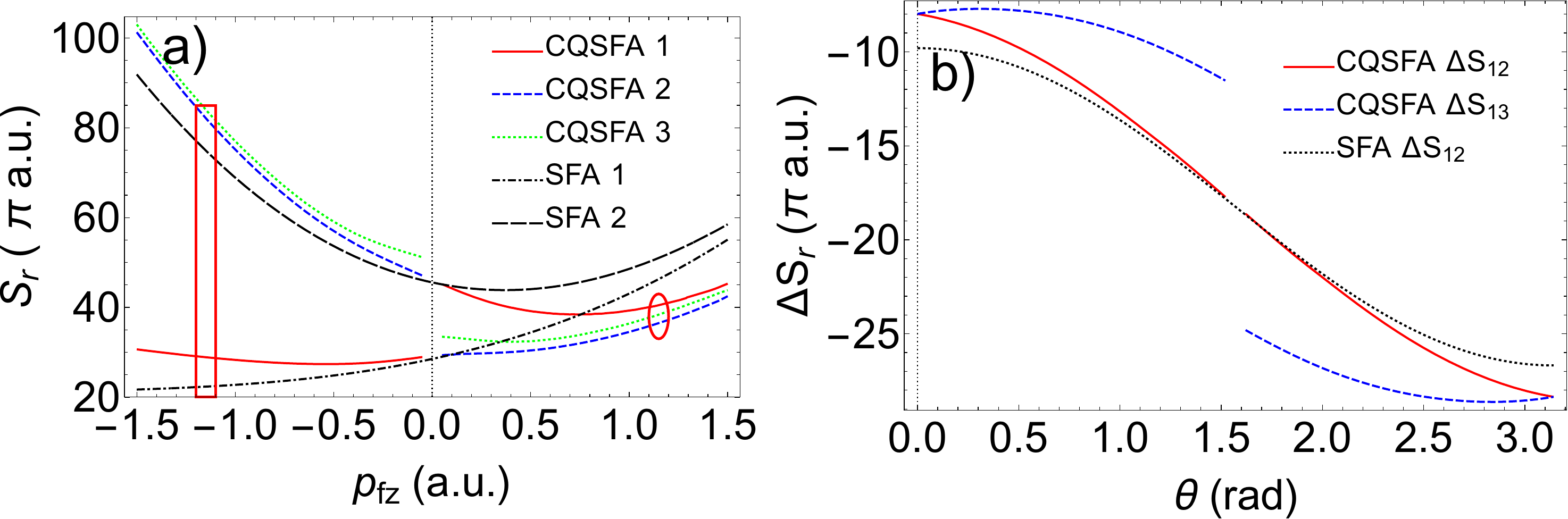}
	\caption{Panel (a) shows the real part of the action for all CQSFA orbits, together with their SFA counterparts, plotted as functions of the final momentum $p_{fz}$, computed for perpendicular final momentum of $p_{fx}=0.25$ a.u.  As in Fig.~\ref{fig:MomReTimes}, type A and B intra-cycle interference is indicated by circles and squares, respectively.  Panel (b) displays the real part of the action differences $\Delta S_{12}$ and $\Delta S_{13}$, together with its SFA counterpart, as functions of the deflection angle $\theta$ and energy $0.1$ a.u.. The remaining parameters are the same as in the previous figures.}
	\label{fig:ReAction}
\end{figure}

\begin{figure}
	\includegraphics[width=0.5\textwidth]{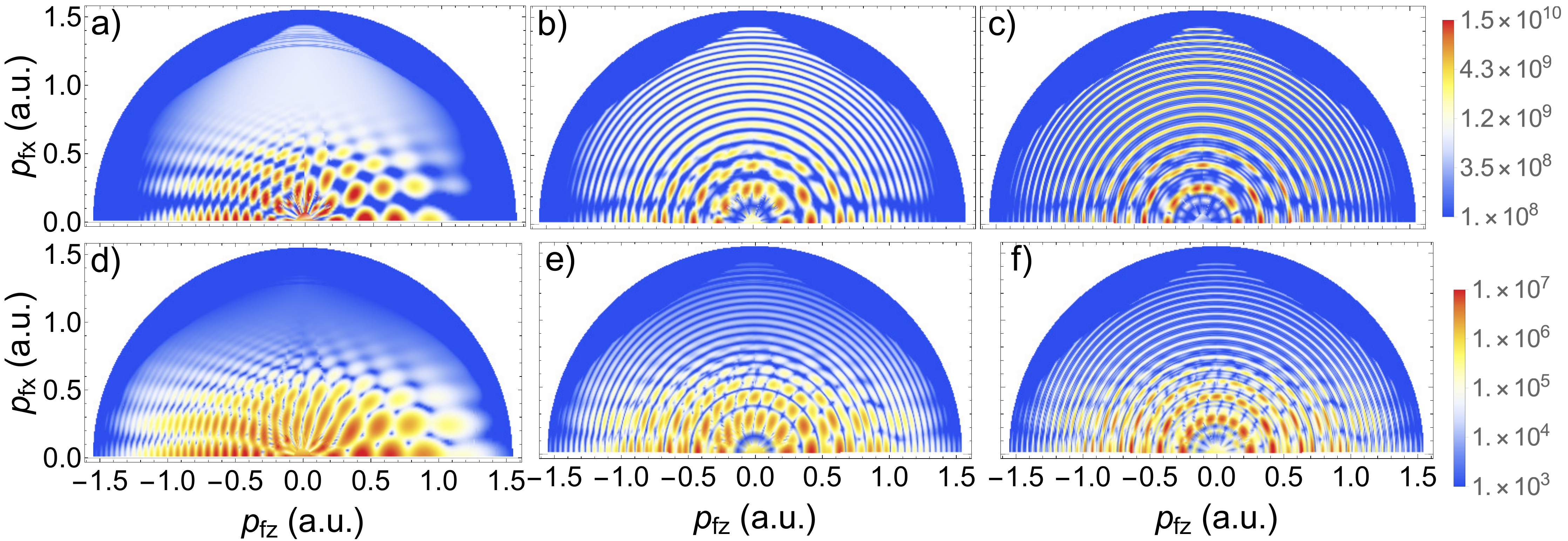}
	\caption{Photoelectron angular distributions computed in the CQSFA using orbits 1, 2 and 3 without symmetrizing with respect to the origin, within one, two and four cycles [left, middle and right panels, respectively]. The upper and lower panels have been computed without and with prefactors, respectively. The field and atomic parameters are the same as in the previous figures. The panels have been plotted in a logarithmic scale.  }
	\label{fig:allcqsfa}
\end{figure}
The real parts of the actions are displayed in Fig.~\ref{fig:ReAction}(a) for the three CQSFA orbits as functions of $p_{fz}$. The figure shows a similar behavior as for the SFA, with type A and B interference corresponding to thicker and finer fringes, respectively. One should note that type A interference is more sensitive to the Coulomb potential, and that, for large positive momentum, the action related to orbit 3 tends to that related orbit 1. This leads to very thick fringes in this momentum region. The real parts of $\Delta S _{ij}$, plotted in Fig.~\ref{fig:ReAction} as a function of the deflection angle, confirm the abovementioned trends. First, the action difference $\Delta S _{12}$ between orbit 1 and 2 tends to the SFA for perpendicular photoelectron emission, but deviates from it for other angles. This causes the  vertical structures in the SFA to be distorted into a fan. In contrast, the difference $\Delta S _{13}$ agrees with its SFA counterpart at the polarization axis, but increases with the scattering angle. This leads to the convergent fringes seen in Fig.~\ref{fig:A_B_ABCQSFA}(d) and  \ref{fig:A_B_ABCQSFApref}(d). In all cases, there is a decrease in $\Delta S _{ij}$ as the polarization axis is approached, which manifests itself as thicker interference fringes. 

If all orbits are considered (Fig.~\ref{fig:allcqsfa}), a more complex pattern arises and several types of fringes are superimposed.  In Figs.~\ref{fig:allcqsfa}(a) and (d), computed within a cycle of the driving field, we see type B and type A intra-cycle interference for negative and positive parallel momentum $p_{fz}$, respectively. Particular visible are the nearly horizontal fringes caused by the interference of type 2 and 3 trajectories, and the structures related to the interference of orbits 1 and 2. This holds both in the presence and in the absence of prefactors, whose main effect is to introduce a bias towards the $p_{fz}$ axis. Traces of the patterns caused by the interference of type 1 and 3 trajectories can also be identified, but they are much less prominent. This is possibly caused by their contrast being poorer than that of the other patterns [see  Figs.~\ref{fig:A_B_ABCQSFA}(f) and \ref{fig:A_B_ABCQSFApref}(f)].
 
If more cycles are included [middle and right columns of Fig.~\ref{fig:allcqsfa}], there will be circular inter-cycle fringes dictated by Eq.~(\ref{eq:nintercycle}), which tend towards a Dirac delta comb as the number of cycles increase. In addition, intra-cycle fringes may be either washed out or reinforced. For instance, the convergent structure due to the interference of orbits 1 and 3 is no longer visible, and the nearly horizontal fringes related to the interference of orbits 2 and 3 is weakened.  In contrast, the fan-shaped structure from the interference of orbits 1 and 2, and the spider-like structure near the origin from the interference of orbits 2 and 3 are very clear, and even seem to reinforce each other.  The patterns become increasingly symmetric as more cycles are included in the computation. This can be seen by comparing Figs.~\ref{fig:allcqsfa}(b) and (c), which has been computed for two cycles,  with  Figs.~\ref{fig:allcqsfa}(e) and (f), for which four cycles have been incorporated. 
 
\section{Conclusions}
\label{sec:conclusions}

Using the Coulomb quantum-orbit strong-field approximation (CQSFA) \cite{Lai2015}, we have isolated many types of interference patterns and other qualitative features present in ATI momentum distributions. Apart from the widely studied near-threshold fan-shaped structure, the inter-cycle ATI interference rings and the ATI side lobes, these features include many types of intra-cycle interference that have been overlooked in the literature. We provide direct evidence of how these patterns form, and show that they may be viewed as holographic type structures arising from different types of interfering trajectories.  We follow the notation in  \cite{Yan2010,Yan2012,Lai2015,Lai2017}, which classifies the trajectories that reach the detector directly as type 1 orbits, and those that leave from the opposite side and are deflected by the core without undergoing hard collisions as type 2 and 3 orbits. Previously overlooked holographic patterns  that have been studied in this work include finer  structures that arise from the intra-cycle interference of events separated by more than half a cycle, nearly horizontal, broad interference fringes stemming from the interference of type 2 and 3 trajectories, and a converging structure caused by the interference of type 1 and 3 trajectories. Within many field cycles, some of these structures may be weakened, washed out or reinforced. 

We have found that orbit 3 is pivotal for many ATI features and have provided a systematic analysis of its effects. In previous studies \cite{Yan2012,Lai2017} this orbit has been neglected, possibly because the corresponding prefactor  strongly reduces the overall signal. Our studies show that, outside the $p_{fz}$ axis, this counteracts the fact that ionization probability along this orbit is quite high. However, two peaks remain located on the $p_{fz}$ axis which contribute to the sidelobes identified in \cite{Huismanset2010Science}. Interference between orbits 2 and 3 produces a spider-like interference pattern, which can be seen superimposed on the finger-like interference pattern that occurs due to interference between orbits 1 and 2. The same spider-like pattern is seen in \cite{Xie2016,LiPRL2014}, in which the quantum-trajectory Monte Carlo (QTMC) model is applied to mid-IR fields, and experimentally in \cite{Hickstein2012,Bian2011PRA}, and it is attributed to these forward scattered trajectories. The on-axis contribution of orbit 3 to the overall PADs improves the agreement with the time-dependent  Schr\"odinger equation (TDSE) \cite{Yan2012,Lai2017, Bian2011PRA} and with experiments \cite{Rudenko2004JPB,Maharjan2006JPB, Hickstein2012,Haertelt2016PRL}, and can be seen in Coulomb-corrected computations in which orbit 3 has been included implicitly \cite{Shvetsov-ShilovskiPRA2016,Xie2016}.  It is also worth noting that classical soft forward-scattered trajectories associated with the low-energy structure (LES) \cite{Kaestner2012JPB} are the same type of trajectories as orbit 3.  This correspondence is possible because we are solving Newton's equations of motion for the continuum. Hence, all our orbits in the continuum have direct classical counterparts. Thus, classical or quasi-classical methods may be built from the CQSFA by performing incoherent sums over trajectories, neglecting or approximating prefactors, and ignoring sub-barrier corrections. For other types of trajectories see our previous publications \cite{Wu2013a,Wu2013b,Zagoya2014,Symonds2015}

We also derive conditions for interference patterns, which are kept as general as possible with regard to the number of field cycles and events per cycle, and provide an analytic expression determining the overall shapes of the distributions. Using 
properties related to the field being monochromatic, we show analytically that the intercycle interference condition is the same for both the SFA and CQSFA. The shape of the distributions and other features will however be affected by the Coulomb potential. We also provide a more rigorous discussion of the sidelobes than what currently exists in the literature, and show that they are mainly determined by the behavior of orbits 2 and 3. In particular, the imaginary part of the action mirrors the behavior of those of the ionization times $t_{2c}$ and $t_{3c}$, which exhibit minima for non-vanishing parallel momenta. This is both verified numerically and analytically using the long-wavelength approximation. The sub-barrier integral over the binding potential is also computed analytically, and is shown to exert a strong influence on the shapes of the PADs.

Furthermore, we make a detailed assessment of intra-cycle interference, and the quantum-orbit analysis in this work strongly suggests that the conditions derived in \cite{Yan2012} are only valid for high momenta and orbits 1 and 2. This is because, in \cite{Yan2012}, the imaginary parts of the times related to orbits 1 and 2 are set to be equal and their momenta at the tunnel exit is chosen to be equal to their final momenta.  These assumptions hold in the SFA and are good approximations for high momenta, as fast electrons are less influenced by the Coulomb potential. This is consistent with our analysis, which shows that the initial momenta $\mathbf{p}^{(1)}_0$, $\mathbf{p}^{(2)}_a$ and the ionization times $t_{1c}$ and $t_{2c}$ tend to their SFA counterparts in this regime. For momenta close to the threshold, however, these assumptions no longer hold. Additionally, one should be careful considering interference between orbit 1 and 2 when $p_{x 0}=0$, as orbit 2 cannot have zero initial perpendicular momenta or it will  undergo a hard collision with the parent ion. For orbit 3, the conditions in \cite{Yan2012} are not applicable in any momentum range, as it behaves in a very different way. Apart from having a much larger tunnel probability throughout, which implies a much smaller $\mathrm{Im}[t_{3c}]$, it does not tend to   the SFA as the momentum increases. Luckily, the prefactor suppresses this orbit over a wide range of scattering angles. However, this is not the case near the field-polarization axis. 
 
This analysis is greatly facilitated by how the CQSFA is implemented. While our method is similar to other approaches such as the trajectory-based Coulomb-corrected strong-field approximation (TCSFA) \cite{Yan2010,Yan2012} and the QMTC model  \cite{LiPRL2014,Xie2016,Shvetsov-ShilovskiPRA2016}, there are some key differences.  The TCSFA and the QMTC method solve the forward problem. Since it is not known what the final momentum will be given a particular starting momentum, one must use larger initial momentum regions in order to sample the final momentum region of interest. Thus, many trajectories, typically $10^8-10^9$, with different initial momenta must be run before each bin is sufficiently populated and interference patterns can be resolved. Furthermore, a uniform spread of initial momenta may undersample specific types of trajectories. This is particularly true for type 3 orbits, whose initial momenta are strongly bunched close to the parallel momentum axis. In contrast, the CQSFA solves the inverse problem, so that for each point in final momentum there are three well defined orbits and we only need to calculate as many points as the resolution dictates.

In addition, the TCSFA has caustics that are made worse both by including sub-barrier Coulomb corrections and when orbit 3 or 4 are included, and which make the interference patterns less clear \cite{Yan2012}. This could be due to orbit 3 becoming more chaotic for low momenta, which may be problematic if a forward mapping is performed. The CQSFA does not suffer from this despite considering sub-barrier Coulomb corrections and orbit 3. The QTMC methods do not contain caustics but are less general as they either disregard sub-barrier corrections \cite{Shvetsov-ShilovskiPRA2016} or they use quasi-static tunnelling rates \cite{LiPRL2014, Xie2016}, which will not be valid for higher frequencies \cite{Lai2017}.

In the CQSFA, the momentum is approximated to be constant in the sub-barrier part of the contour, as originally done in \cite{Yan2010,Yan2012}. One of the main issues with this region is that one must integrate the potential up its singularity. In practice, one must determine a lower bound for which there are no qualitative changes in the PADs. However, this introduces some ambiguity so that no quantitative statements can be made about total ionization rates. Furthermore, the tunnel trajectory end point is fixed by the tunnel exit derived in \cite{PPT1967}, which takes a perturbative approach. Improvements to this contour have been reported in \cite{Pisanty2016b} in the context of low-energy structures.
Nonetheless, the CQSFA can qualitatively reproduce many features in the ATI momentum distribution, including the number of nodes on each ATI ring. In \cite{Yan2012} it is stated how sub-barrier corrections correct the number of nodes on the second ring but those on the first remain incorrect. This is attributed to tunnel contour approximation being insufficient. However, we do obtain the correct number of fringes using the same approximation. It is more likely that this discrepancy is due the $\dot{\bm{p}}\cdot\bm{r}$ term used in our expression, which is absent in \cite{Yan2012}. In previous publications, we have found this term to be  important for a good agreement with the TDSE \cite{Lai2015,Lai2017}. Similar results have also been reported in \cite{Shvetsov-ShilovskiPRA2016}.  Given that the CQSFA has very low computational demands, it can be extended to more complex systems such as multi-electron targets.

\acknowledgements
We thank X. Y. Lai and L. B. Madsen for useful discussions, and the UK Engineering and Physical Sciences Research Council (EPSRC) (grant EP/J019240/1) for financial support. The authors acknowledge the use of the
UCL Legion High- Performance Computing Facility (Legion@UCL), and associated
support services, in the completion of this work.
\appendix
\section{Generalized interference conditions}
\label{app:generalized}
In this appendix we derive Eqs.~(\ref{eq: generalizedinterf}) and (\ref{eq:nintercycle}) from Eq.~(\ref{eq:generalinterf}). First, Eq.~(\ref{eq:generalinterf}) is rewritten as
\begin{align}
 \Omega(\mathbf{p}_f)&=\left|\sum_{e=1}^{n_{e}}
 \sum_{c=0}^{N_c-1}\exp[iS_{ec}]\right|^2\\
 &=\left|\sum_{e=1}^{n_{e}}\exp[iS_{e0}]
 \sum_{c=0}^{N_c-1}\exp[i(S_{ec}-S_{e0})]\right|^2.
\end{align}
From Eq.~(\ref{eq:intercycle}) we can calculate $S_{ec}-S_{e0}$, which reads as 
\begin{align}
	S_{ec}-S_{e0}=\frac{2\pi i c}{\omega}\underbrace{\left(\ip+\up+\frac{1}{2}\bm{p}_f^2\right)}_{\alpha}.
\end{align}
The fact that we  can pull out a factor $S_{e0}$ and the remaining sum over $c$ is not dependent on $e$ means that we can factorise the two sums. This gives
\begin{align}
	\Omega(\mathbf{p}_f)&=\underbrace{\left|\sum_{e=1}^{n_{e}}\exp[iS_{e0}]\right|^2}_{\Omega_{n_e}}
	\underbrace{\left|\sum_{c=0}^{N_c-1}\exp\left[\frac{2\pi i\alpha c}{\omega}\right]\right|^2}_{\Omega_{N_c}},
	\intertext{from which we can infer Eq.~(\ref{eq: generalizedinterf}), namely}
	\Omega(\mathbf{p}_f)&=\Omega_{n_e}(\mathbf{p}_f)\Omega_{N_c}(\mathbf{p}_f).
\end{align}
We can further simplify $\Omega_{N_c}(\mathbf{p}_f)$, so that 
\begin{align}
	\Omega_{N_c}(\mathbf{p}_f)
	&=\left|\frac{\exp\left[\frac{2\pi i\alpha (N_c-1)}{\omega}\right]-1}{\exp\left[\frac{2\pi i\alpha}{\omega}\right]-1}\right|^2 \notag\\
	&=\frac{\cos\left[\frac{2\pi i N_c}{\omega}\alpha\right]-1}{\cos\left[\frac{2\pi i }{\omega}\alpha\right]-1},
\end{align}
which leads to Eq.~(\ref{eq:nintercycle}).
\section{Coulomb correction for tunnel prefactor}
\label{app:integral1}
In this appendix, we compute the integral over the binding potential for the imaginary part of the CQSFA action related to tunnel ionization, Eq.~(\ref{eq:ImCQSFAAction}), in the long wavelength approximation. This integral is important in determining the shapes of single-orbit distributions, and influences their sidelobes. 
The tunnel trajectory can be written explicitly as
\begin{equation}
\bm{r}_{0}(\tau)=i \bm{p}_{e0}(\tau_i-t'_{i})+i \int_{t'_{i}}^{\tau_i}\bm{A}(t_{r}+i\tau'_i)d\tau'_i
\end{equation}
Using the long wavelength approximation and expanding around the imaginary component, the above-stated expression is approximated by 
\begin{equation}
\bm{r}_{0}(\tau)=(\tau_i-t'_{i})\left[i(\bm{p}_{0}+\bm{A}(t'_{r}))-\frac{1}{2}\dot{\bm{A}}(t'_{r})(\tau_i+t'_{i})\right],
\end{equation} where $\tau_i= \mathrm{Im}[\tau]$.  This expression can be used to compute the indefinite integral

 {
	\compeqn
	\begin{align}
 	\int V(\bm{r}_{0}(\tau))\mathrm{d}\tau &= \frac{i C}{\sqrt{-\bm{p}_{0x}^{2}+\chi^2}}\Big(\ln(\tau_i-t'_i)  \nonumber \\
 	& \hspace{-1.75cm} -\ln\left(2[\chi\eta(\tau_i)-\bm{p}^2_{0x}]+2\sqrt{-\bm{p}^2_{0x}+\eta(\tau_i)^2}\sqrt{-\bm{p}^2_{0x}+\chi^2}\right)  \Big)
 	\end{align}
 }where 
\begin{eqnarray}
\chi&=&i ({p}_{0x}+A(t'_{r}))-t'_{i} \dot{A}(t'_{r}) \qquad \mathrm{and}\\  
\eta(\tau_i)&=&i ({p}_{0z}+{A}(t'_{r}))-\frac{1}{2}(t'_{i}+\tau_i) \dot{A}(t'_{r}).
\end{eqnarray}
We are however interested in the definite integral from $t'$ to $t'_{r}$. Care must be taken with the lower bound as it will lead to a divergence. For that reason, we take it as $t'-i\Delta \tau_i$, where $\Delta \tau_i$ is chosen to be arbitrarily small. This gives
\begin{widetext}
\begin{equation}
\mathcal{I}_{V_T} =	\int^{t'_{r}}_{t'-i\Delta \tau_i}\hspace*{-0.5cm} V(\bm{r}_0(\tau))\mathrm{d}\tau= i \ln \left[\left(\frac{t'_i \left(\chi \eta(t'_i-\Delta \tau_i) -p^2_{0x}+\sqrt{-\bm{p}^2_{0x}+\eta_(t'_i-\Delta \tau_i)^2}\sqrt{-\bm{p}^2_{0x}+\chi^2}\right)}{\Delta \tau_i \left(\chi \eta(0) -p^2_{0x}+\sqrt{-\bm{p}^2_{0x}+\eta_(0)^2}\sqrt{-\bm{p}^2_{0x}+\chi^2}\right)}\right)^{C/\sqrt{-\bm{p}^2_{0x}+\chi^2}}\right],
\label{eq:Vintanalytic}
\end{equation}
\end{widetext}
so that $\exp [-i\mathcal{I}_{V_T}]$ will be a power of $ C/\sqrt{-\bm{p}^2_{0x}+\chi^2}$ and  $\Delta \tau_i^{-C/\sqrt{-\bm{p}^2_{0x}+\chi^2}}$ will contribute as an orbit independent overall factor multiplying the whole transition amplitude. There is also some freedom on how to approach this limit, and a convenient parametrization, such as $\Delta \tau_i \sim \delta ^{C/\sqrt{-\bm{p}^2_{0x}+\chi^2}}$, may be employed.  Eq.~(\ref{eq:Vintanalytic}) agrees with numerical computations, in which $\Delta \tau_i$ is set to be small.

\end{document}